\documentclass[conference,compsoc]{IEEEtran}
\IEEEoverridecommandlockouts

\usepackage{graphicx}
\usepackage{multirow}
\usepackage{xcolor}
\usepackage[table]{xcolor}
\usepackage{tikz}
\usetikzlibrary{arrows.meta,positioning,fit,backgrounds,calc,patterns,shapes.geometric}
\usepackage{fontawesome5}
\usepackage[scaled=0.92]{helvet}
\usepackage{booktabs}
\usepackage{array}
\usepackage{adjustbox}
\usepackage{tabularx}
\usepackage{amsmath, amssymb}
\usepackage{makecell}
\usepackage{pifont}
\usepackage{xspace}
\usepackage{enumitem}

\newcommand{\pp}{\,pp}
\newcommand{\Latch}{{\textsc{Chasm}}\xspace}
\definecolor{shadeatk}{HTML}{D6604D}  % table shading: significant change toward the attacker
\definecolor{shadedef}{HTML}{4393C3}  % ... and away from it

\newcommand{\ASR}{\mathrm{ASR}}

\newcommand{\Src}{\mathcal{S}}
\newcommand{\Map}{\mathcal{M}}
\newcommand{\Tgt}{\mathcal{T}}
\newcommand{\Tnat}{\ensuremath{\Tgt(x)}}
\newcommand{\Ttr}{\ensuremath{\Tgt(\Map(\Src(x)))}}

\usepackage[most]{tcolorbox}
\usepackage{ragged2e}
\tcbuselibrary{raster,skins}

\newtcolorbox{qualemail}[3][]{
  enhanced,
  colback=white,
  colframe=#3,
  colbacktitle=#3,
  coltitle=white,
  title={#2},
  fonttitle=\bfseries\footnotesize,
  boxrule=0.65pt,
  arc=1.2mm,
  left=1.7mm,
  right=1.7mm,
  top=1.2mm,
  bottom=1.2mm,
  before upper={\justifying\setlength{\parindent}{0pt}\setlength{\parskip}{2pt}},
  #1
}

\usepackage{hyperref}

\usepackage{soul}
\colorlet{chggrey}{black!12}
\newcommand{\chg}[1]{#1}
\ifCLASSOPTIONcompsoc
  \usepackage[nocompress]{cite}
\else
  \usepackage{cite}
\fi
\ifCLASSINFOpdf
\else
\fi
\begin{document}
% \title{Latent-in-the-Middle: On the Security of Cross-Model Latent Communication in Multi-Agent Systems}
\title{Latent-in-the-Middle: On the Security of Cross-Model 
Latent Communication in Multi-Agent Systems}

\author{%
\IEEEauthorblockN{%
  Manit Baser\textsuperscript{1*} \quad
  Vyoman Jain\textsuperscript{2*} \quad
  Rachel Yeo Hui Min\textsuperscript{1} \\
  Xianglin Yang\textsuperscript{1} \quad
  Meenatchi Sundaram Muthu Selva Annamalai\textsuperscript{3} \quad
  Dinil Mon Divakaran\textsuperscript{4}
}
\IEEEauthorblockA{%
  \textsuperscript{1}National University of Singapore \\
  \textsuperscript{2}Singapore University of Technology and Design \\
  \textsuperscript{3}University of Toronto \\
  \textsuperscript{4}A*STAR Institute of Advanced Intelligence and Computing (A*STAR IAIC)
}
\thanks{* Equal contribution}
}

\maketitle

\begin{abstract}

As large language models increasingly collaborate in heterogeneous multi-agent systems, latent communication has emerged as an efficient alternative to exchanging natural-language messages. By sharing internal representations, agents can avoid repeatedly processing the same context, reducing communication overhead and prefill computation. One such mechanism is cross-model Key–Value (KV) cache transfer, which uses a learned mapper to translate a source model's KV cache into one a target model can use. However, this handoff means the target model consumes context it did not encode itself, thus calling into question whether its safety alignment still holds.

In this work, we investigate the security implications of latent handoffs by introducing \chg{{\Latch}}, a framework characterizing cross-model handoff attacks across three new attack surfaces: the source model, mapper, and handoff mechanism itself. Our experiments reveal that latent handoffs substantially undermine safety alignment and prompt-injection defenses. Notably, a target model that follows only 0.4\% of benchmark prompt injections natively, follows \chg{up to 44.2\%} when transferred through a mapper from a source model. On the other hand, for jailbreak attacks, transfer alone raises attack success by up to 24.6 pp \chg{ within a model family and zero-shot attacks by up to 69.2 pp}, whereas white-box optimization \chg{through the mapper} boosts the attack success rate to 92.5\%\chg{, against 37.5\% for ordinary GCG on the same target}. To address these vulnerabilities, we propose a lightweight target-side cache guard that \chg{corrects transferred caches to restore the target model's refusal}, \chg{considerably} mitigating these attacks.

\end{abstract}


% \fi
\IEEEpeerreviewmaketitle
% \begin{figure*}[t]
% \centering
% \resizebox{\textwidth}{!}{\input{Figures/overview}}
% \caption{Overview. \textcircled{\scriptsize 1} Natively, the aligned target reads the prompt itself. \textcircled{\scriptsize 2} Under cross-model transfer, a source model prefills all but the last token, a mapper fitted on calibration data converts the source cache, and the target decodes from the mapped cache. \textcircled{\scriptsize 3} Every component transfer adds is an entry point for an attacker (red badges); we design one attack class per entry point.}
% \label{fig:overview}
% \end{figure*}

\begin{figure*}[t]
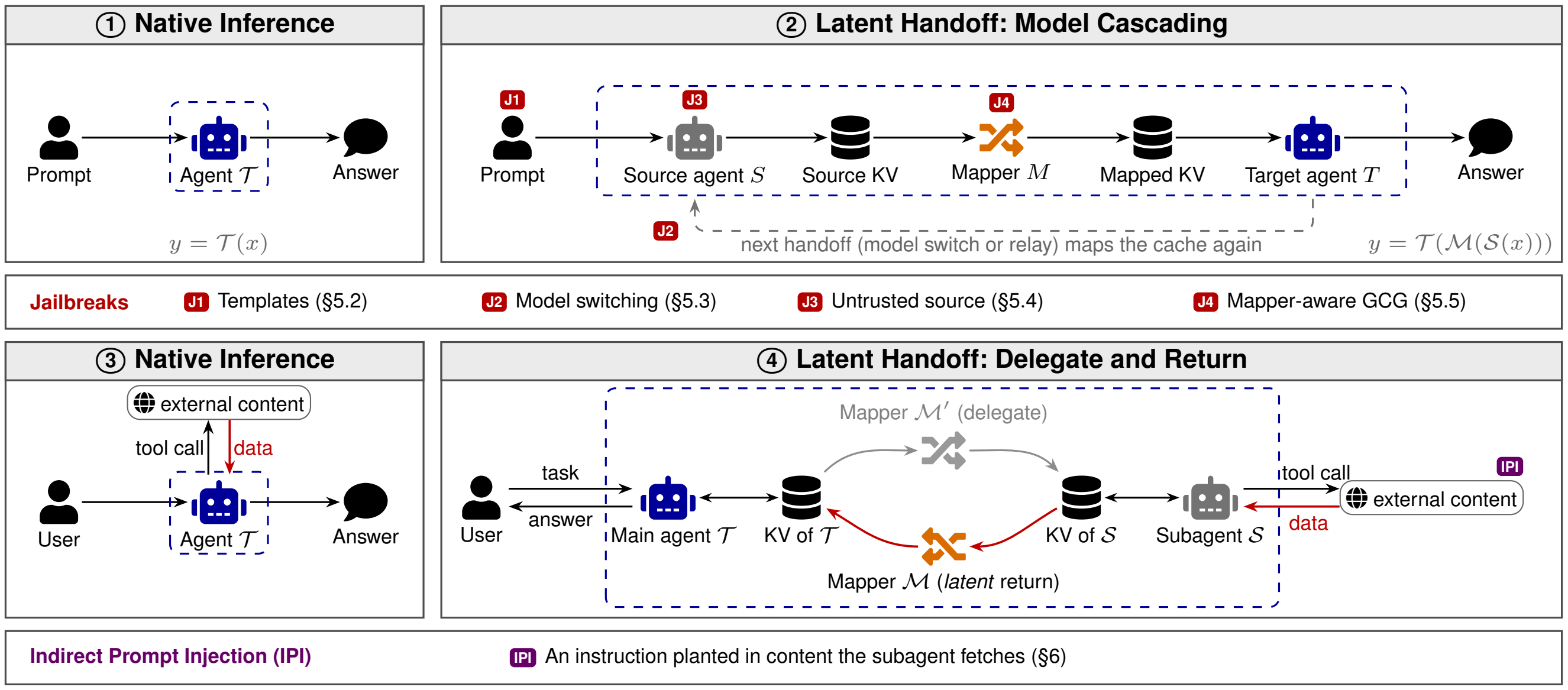

\centering
\resizebox{0.98\textwidth}{!}{%
% \resizebox{0.95\textwidth}{!}{%
\begin{tikzpicture}
\input{Figures/styles}
% ================================================================ row A: jailbreaks
% ---- (1) native, prompt only
\begin{scope}[shift={(0,4.75)}]
\draw[panel] (0,0) rectangle (4.7,2.88);
\draw[head] (0,2.45) rectangle (4.7,2.88) node[midway] {\textcircled{\scriptsize 1} Native Inference};
\node[ico] (nu) at (0.6,1.4) {\faUser};
\node[lab, below=0pt of nu] {Prompt};
\node[ico, text=blue!60!black] (nt) at (2.4,1.4) {\faRobot};
\node[lab, below=0pt of nt] {Agent $\Tgt$};
\node[ico] (nr) at (4.05,1.4) {\faComment};
\node[lab, below=0pt of nr] {Answer};
\draw[flow] (nu) -- (nt); \draw[flow] (nt) -- (nr);
\draw[trust] (1.85,0.8) rectangle (2.95,1.8);
\node[lab, text=black!60, anchor=south] at (2.4,0.04) {$y = \Tgt(x)$};
\end{scope}
% ---- (2) model cascading: one agent encodes, another answers
\def\hx{4.9}\def\hy{4.75}\def\handoffbadges{main}
\def\handofftitle{\textcircled{\scriptsize 2} Latent Handoff: Model Cascading}
\input{Figures/handoff_panel}
% ---- legend A
\begin{scope}[shift={(0,4.0)}]
\draw[panel] (0,0) rectangle (17.5,0.6);
\node[font=\sffamily\scriptsize\bfseries, text=red!70!black, anchor=west] at (0.15,0.3) {Jailbreaks};
\foreach \x/\b/\t/\s in {2.0/J1/Templates/sec:jb:templates, 5.35/J2/Model switching/sec:jb:switching, 8.9/J3/Untrusted source/sec:jb:swap, 13.35/J4/Mapper-aware GCG/sec:jb:whitebox} {
  \node[jb, anchor=west] (b) at (\x,0.3) {\b};
  \node[lab, anchor=west] at ($(b.east)+(0.05,0)$) {\t{} (\S\ref{\s})};
}
\end{scope}
% ================================================================ row B: prompt injection
% ---- (3) native, the agent fetches the content itself
\begin{scope}[shift={(0,0.75)}]
\draw[panel] (0,0) rectangle (4.7,3.1);
\draw[head] (0,2.67) rectangle (4.7,3.1) node[midway] {\textcircled{\scriptsize 3} Native Inference};
\node[ico] (mu) at (0.6,1.3) {\faUser};
\node[lab, below=0pt of mu] {User};
\node[ico, text=blue!60!black] (mt) at (2.4,1.3) {\faRobot};
\node[lab, below=0pt of mt] {Agent $\Tgt$};
\node[ico] (mr) at (4.05,1.3) {\faComment};
\node[lab, below=0pt of mr] {Answer};
\node[cloud] (mext) at (2.4,2.42) {\faGlobe\ external content};
\draw[flow] ($(mt.north)+(-0.12,0.02)$) -- node[lab, left] {tool call} ($(mext.south)+(-0.12,0)$);
\draw[bad] ($(mext.south)+(0.12,0)$) -- node[lab, right, text=red!75!black] {data} ($(mt.north)+(0.12,0.02)$);
\draw[flow] (mu) -- (mt); \draw[flow] (mt) -- (mr);
\draw[trust] (1.85,0.72) rectangle (2.95,1.64);
\end{scope}
% ---- (4) delegate and return
\def\hx{4.9}\def\hy{0.75}
% Delegate-and-return drawing for Fig. 1 (prompt-injection row).
% Origin: bottom-left corner (\hx,\hy); width 12.6, height 3.1.
% Each agent keeps one KV cache. Delegation (gray, top): KV of T through M'
% into KV of S. Return (red, bottom): KV of S through M into KV of T.
\begin{scope}[shift={(\hx,\hy)}]
\draw[panel] (0,0) rectangle (12.6,3.1);
\draw[head] (0,2.67) rectangle (12.6,3.1) node[midway] {\textcircled{\scriptsize 4} Latent Handoff: Delegate and Return};
% user and main agent
\node[ico] (u) at (0.45,1.35) {\faUser};
\node[lab, below=0pt of u] {User};
\node[ico, text=blue!60!black] (t) at (2.55,1.35) {\faRobot};
\node[lab, below=0pt of t] {~~Main agent $\Tgt$};
\draw[flow] ($(u.east)+(0.05,0.1)$) -- node[lab, above=1.5pt, pos=0.42] {task} ($(t.west)+(-0.05,0.1)$);
\draw[flow] ($(t.west)+(-0.05,-0.1)$) -- node[lab, below=1.5pt, pos=0.58] {answer} ($(u.east)+(0.05,-0.1)$);
% the two caches
\node[ico] (kT) at (4.05,1.35) {\faDatabase};
\node[lab, below=0pt of kT] {KV of $\Tgt$};
\node[ico] (kS) at (7.2,1.35) {\faDatabase};
\node[lab, below=0pt of kS] {KV of $\Src$};
\draw[<->, >={Stealth[length=4pt]}, line width=0.6pt] (t) -- (kT);
% subagent and external content
\node[ico, text=black!55] (s) at (8.65,1.35) {\faRobot};
\node[lab, below=0pt of s] {Subagent $\Src$};
\draw[<->, >={Stealth[length=4pt]}, line width=0.6pt] (kS) -- (s);
\node[cloud, anchor=west] (ext) at (10.1,1.35) {\faGlobe\ external content};
\draw[flow] ($(s.east)+(0.02,0.1)$) -- node[lab, above=1.5pt, pos=0.76] {tool call} ($(ext.west)+(0,0.1)$);
\draw[bad] ($(ext.west)+(0,-0.1)$) -- node[lab, below=1.5pt, pos=0.32, text=red!75!black] {data} ($(s.east)+(0.02,-0.1)$);
% delegation: KV of T -> M' -> KV of S (top)
\node[ico, text=black!40] (mp) at (5.65,1.88) {\faRandom};
\node[lab, above=0pt of mp, text=black!50] {Mapper $\Map'$ (delegate)};
\draw[flow, black!45] (kT.north east) to[out=45, in=180] (mp.west);
\draw[flow, black!45] (mp.east) to[out=0, in=135] (kS.north west);
% return: KV of S -> M -> KV of T (bottom)
\node[ico, text=orange!85!black] (m) at (5.65,0.8) {\scalebox{-1}[1]{\faRandom}}; % mirrored: the return flows right to left
\node[lab, below=0pt of m] {Mapper $\Map$ (\textsl{latent} return)};
\draw[bad] ($(kS.west)+(-0.02,-0.12)$) to[out=-150, in=0] (m.east);
\draw[bad] (m.west) to[out=180, in=-30] ($(kT.east)+(0.02,-0.12)$);
% components on the path from input to answer
\draw[trust] (1.85,0.12) rectangle (9.42,2.58);
% badge
\node[pi, anchor=south east] at ($(ext.north east)+(0,0.04)$) {IPI};
\end{scope}
% ---- legend B
\begin{scope}[shift={(0,0)}]
\draw[panel] (0,0) rectangle (17.5,0.6);
\node[font=\sffamily\scriptsize\bfseries, text=violet!80!black, anchor=west] at (0.15,0.3) {Indirect Prompt Injection (IPI)};
\node[lab, anchor=center] (ipitext) at (9.0,0.3) {An instruction planted in content the subagent fetches (\S\ref{sec:injection})};
\node[pi, anchor=east] at ($(ipitext.west)+(-0.05,0)$) {IPI};
\end{scope}
\end{tikzpicture}}
\caption{Overview of native inference and latent handoffs. \emph{Top: jailbreaks.} \textcircled{\scriptsize 1} Under native inference, the aligned agent $\Tgt$ encodes the prompt and generates the answer. \textcircled{\scriptsize 2} Under model cascading, source agent $\Src$ encodes the prompt, and mapper $\Map$ converts its KV cache for target agent $\Tgt$. Subsequent model switches or relays trigger further mappings. \emph{Bottom: indirect prompt injection (IPI).} \textcircled{\scriptsize 3} Under native inference, the agent fetches and encodes external content. \textcircled{\scriptsize 4} Under delegate and return, the main agent transfers its cache to subagent $\Src$ via mapper $\Map'$, then receives the subagent's cache via mapper $\Map$, answering without encoding the fetched content itself. Red arrows denote untrusted data. Red badges mark the four jailbreak entry points: prompt templates (J1), model switching (J2), an untrusted source (J3), and mapper-aware GCG (J4). The violet badge marks an instruction planted in content fetched by the subagent. Dashed boxes enclose the computation between input and answer.}
\label{fig:overview}
\end{figure*}

\section{Introduction}
\label{sec:intro}

Large language models (LLMs) increasingly operate as heterogeneous agents that delegate tasks, exchange intermediate results, and coordinate to solve problems beyond the reach of a single model. This shift is underway: a 2026 KPMG survey reports that 25\% of large U.S. enterprises are developing multi-agent systems~\cite{KPMG2026Q3AIPulse}, while Gartner expects that more than 60\% of multi-agent systems will combine agents from different vendors by 2028~\cite{gartner2025multiagent}. 
Meanwhile, LLM applications increasingly distribute work across models: systems cascade from cheaper models to larger ones~\cite{chen2024frugalgpt}, route requests to the smallest capable model~\cite{ong2025routellm, jitkrittum2026universal}, or compose agents built on different models~\cite{fu2026cachetocache}. 
At the same time, agents are moving from private chats into shared workspaces such as Slack, Discord, and group chats, where persistent agents observe messages from multiple users as context~\cite{patlan2026crossuser}. 
Across these deployments, context routinely crosses between models and trust boundaries.
% Managed services and open-source serving stacks increasingly support routing across model backends~\cite{aws2025bedrockrouting,microsoft2026modelrouter,nvidia2026switchyard,vllm2026productionstack}. 

However, passing context across agents can be communication-intensive.
In SwarmChat~\cite{aggarwal2026swarmchat}, eight agents exchanged 2,482 messages during a six-hour collaborative coding run. Agents commonly communicate through natural-language messages or shared memory, requiring target agents to repeatedly process context that may already have been processed elsewhere. 
This prefill computation accumulates over a long interaction. Anthropic reports that multi-agent implementations consume 3--10$\times$ more tokens than single-agent approaches on comparable tasks~\cite{phillips2026building}. Beyond cost, natural-language communication can introduce ambiguity and information loss, contributing to coordination failures~\cite{du-etal-2026-enabling, cemri2026why}.

These efficiency limitations have motivated \emph{latent-space communication}, in which agents exchange internal states rather than natural-language messages. 
One such reusable latent state is the Key--Value (KV) cache. 
Production serving systems already support KV-cache transfer and reuse across hosts, requests, and storage tiers through disaggregated inference and cache offloading~\cite{jia2025disaggregated,vllm2026kvoffload, 3724648.3724658}. 
However, because a KV cache depends on a model's specific architecture and weights, it generally is not directly transferable across models.
To address this incompatibility, recent work introduces cross-model KV-cache transfer via learned mappers~\cite{qu2026cachebridgeefficientcrossmodelkv,alabdulkareem2026swarm,heo2026closedform,castin2026kv}. 
For example, NVIDIA proposes a closed-form, per-head ridge mapper that transfers caches across differently sized models within the same model family, retaining 73--98\% of target-model accuracy while reducing prefill latency by 2.7--25$\times$~\cite{heo2026closedform}. 
Subsequent work explores MLP and transformer-based mappers~\cite{heo2026closedform,alabdulkareem2026swarm}, head-local variants~\cite{qu2026cachebridgeefficientcrossmodelkv}, and distillation-trained linear translators~\cite{castin2026kv}, and we are starting to see mapped prefill offered as a hosted service~\cite{onetriangle2026cachetransfer}.
Across these designs, a latent handoff proceeds in three steps:
a \emph{source} model prefills the context into its KV cache; 
a \emph{mapper}, fitted or trained offline, translates that cache into the representation space of the \emph{target} model;
and the target model processes only the remaining suffix before generating its answer directly from the converted cache.

% try to make the flow more smooth by adding a sentence
In practice, this three-step handoff also crosses trust and safety boundaries as illustrated in Fig.~\ref{fig:overview}.
Consider a coding assistant deployed in a team's workspace, where a user asks it to fix a bug described in an issue that links to a web page.
Its \emph{main agent} is a large model that has been safety-tested and hardened against prompt injection, as coding assistants routinely process untrusted files and pages.
To save cost, the main agent delegates reading the issue and the linked page to a cheaper \emph{subagent}, as coding assistants already do with text~\cite{anthropic2026subagents,openai2026codexsubagents}.
With latent communication, the subagent does not reply with a summary. It fetches the page with a tool call, and its KV cache returns to the main agent through a mapper (Fig.~\ref{fig:overview}~\textcircled{\scriptsize 4}).
The main agent then writes the patch without ever computing its own representation of the page. 
Session-level handoffs create the same separation between who reads the context and who answers: a cheaper model may escalate a hard request to the larger model in a cascade~\cite{heo2026closedform}, or a user may switch models mid-session, likewise transferring the cache through a mapper (Fig.~\ref{fig:overview}~\textcircled{\scriptsize 2}). 
Similar cross-model handoffs arise in model routers~\cite{ong2025routellm}, multi-agent group chats, and shared workspaces~\cite{patlan2026crossuser}. 
In each, the model that answers decodes from a cache that another model computed and a mapper rewrote.

Although cross-model KV-cache transfer aims to preserve accuracy while reducing latency, delegating representation introduces security risks by making an agent's behaviour depend on computation performed elsewhere. 
A model's refusal decision is read from its internal representation of the prompt~\cite{arditi2024refusal}, and the first few answer tokens---where safety alignment does most of its work~\cite{qi2025shallow}---are generated from that representation. 
Prompt-injection defenses likewise rely on representing trusted instructions and untrusted data differently~\cite{chen2025secalign}. 
Natively, the answering model builds these representations itself; in a latent handoff, it inherits a cache computed by another model and rewritten by a mapper. 
In the coding-assistant example above, a jailbreak in the user's request is first encoded by the cheaper session model, and an instruction planted in the linked page is encoded only by the subagent---so the hardened main agent never sees either input in the form it was trained to handle. 
As a result, latent handoffs expand the attack surface for threats such as jailbreaks, where a user seeks harmful answers, and indirect prompt injection (IPI), where a third party plants instructions in external content~\cite{greshake2023not}.

To study these risks systematically, we introduce \textbf{\Latch} (\textbf{C}ross-Model \textbf{H}andoff \textbf{A}ttacks in Latent-\textbf{S}pace Agent Co\textbf{m}munication), which models the components separating the input from the aligned answering agent. 
A handoff interposes a source that encodes the input and a mapper that converts its cache, and a session may chain multiple handoffs. 
Each component creates a point where an attack can succeed even if it fails against the standalone target model, and \Latch organizes attacks by the entry point they exploit (Fig.~\ref{fig:overview}): 
through the \emph{input}, a published jailbreak template (J1) or an instruction in fetched content (IPI) is encoded by a model other than the answering model; 
through \emph{repeated handoffs}, a cache is remapped across successive model switches (J2); 
through the \emph{source}, an attacker-controlled source model computes the cache (J3); 
and through the \emph{mapper}, prompts are optimized with gradients end-to-end through the transfer path (J4). 
All attacks except J4 are \emph{zero-shot}, requiring no training, gradients, or search. 
% We measure each attack as the change in success rate through the handoff relative to native inference on the target model, and study additional handoff-specific entry points with no native counterpart in the appendix.
We measure each attack as the change in success rate through the handoff relative to native inference on the target model, and study additional handoff-specific entry points with no native counterpart in Appendices~\ref{app:privacy} and~\ref{app:backdoor}.

\textbf{Findings and contributions.} With \Latch,~we evaluate three model families (Qwen3, Llama~3, and Ministral~3) and four mapper types, each calibrated on two corpora, for 24 settings. Our contributions and main findings are as follows.

{\setlength{\leftmargini}{1.3em}\setlength{\topsep}{2pt}
\begin{itemize}\setlength{\itemsep}{1pt}\setlength{\parskip}{0pt}
\item[$\rightarrow$] \emph{A framework for handoff attacks}.~\Latch organizes attacks on latent handoffs by the entry point they exploit
%, namely the input, repeated handoffs, the source, and the mapper 
(Section~\ref{sec:threat}).
\item[$\rightarrow$] \emph{Zero-shot jailbreaks get stronger (J1--J3).} Through the handoff, zero-shot attacks raise the HarmBench attack success rate (ASR) by up to +69.2 percentage points (pp) without touching the target model or the deployed mapper, and an untrusted source succeeds in all 24 settings (Section~\ref{sec:jailbreaks}).
\item[$\rightarrow$] \emph{The mapper hands the attacker a stronger white-box attack (J4).} Suffixes optimized through the mapper cross 90\% ASR on every target model through the SwaT (Chat) renderers, whereas ordinary GCG on a standalone target model reaches at most 65.0\% (Section~\ref{sec:jb:whitebox}).
\item[$\rightarrow$] \emph{Injection defenses fail across the handoff (IPI).} In the first evaluation of indirect prompt injection across a latent handoff, the injection ASR of Meta-SecAlign-8B on AlpacaFarm rises from 0.4\% natively to about 30\% through the four default FineWeb-calibrated mappers, and to as much as 44.2\% through SwaT (Chat), when a 3B subagent encodes the content. The pipeline is only as robust as the agent that reads the content (Section~\ref{sec:injection}).
% \item[$\rightarrow$] \emph{An existing defense fails; we propose a new one.} VReparo~\cite{lin2026cache} fails against adaptive attackers and untrusted source models. Our~\mbox{\textsc{HandoffGuard}}, a low-rank correction of the mapped cache, brings attacker-controlled sources within 4.5 points of native HarmBench ASR across all families at no utility loss, though a white-box attacker who fine-tunes the source still defeats it (Section~\ref{sec:defenses})
\item[$\rightarrow$] \emph{A new defense.} 
%VReparo~\cite{lin2026cache} fails against adaptive attackers and untrusted sources. 
We propose~\mbox{\textsc{HandoffGuard}}, a low-rank correction of the mapped cache that brings attacker-controlled sources within 4.5 points of native HarmBench ASR across all families at \chg{negligible} utility loss, though a costly adaptive attack that fine-tunes the source still succeeds (Section~\ref{sec:defenses}).
\end{itemize}}
\noindent Furthermore, we show that a mapped cache leaks the prompt (Appendix~\ref{app:privacy}), that a mapper can be backdoored (Appendix~\ref{app:backdoor}), and that repeated handoffs erode the target model's capability (Appendix~\ref{app:erosion}).

\section{Background}
\label{sec:background}

\subsection{KV Caches and Prefill Reuse}
\label{sec:bg:kv}

A decoder-only LLM processes a token sequence $x = (x_1,\dots,x_n)$ in two phases. 
During \emph{prefill}, each of the $L$ transformer layers computes a key $\mathbf{k}^{\ell,h}_p$ and value $\mathbf{v}^{\ell,h}_p$ for every position $p$ and KV head $h$, forming the KV cache
\[
\mathcal{C}(x) = \{(\mathsf{K}^\ell,\mathsf{V}^\ell)\}_{\ell=1}^{L},
\]
where $\mathsf{K}^\ell$ and $\mathsf{V}^\ell$ aggregate keys and values across positions and heads (with rotary position embeddings, RoPE, applied to keys but not values). 
During \emph{decode}, each newly generated token attends to cached states and appends its own keys and values. 
Since prefill cost grows with prompt length, serving systems reuse, offload, and move caches across requests and hosts. 
Cache reuse, however, requires the consumer to be the same model that produces it.

\subsection{Cross-Model Latent Communication}
\label{sec:bg:transfer}

To share context across heterogeneous models without re-prefilling, a \emph{mapper} $\Map$ translates a source model $\Src$'s KV cache into the representation space of a target model $\Tgt$:
\[
\Map:\; \mathcal{C}_{\Src}(x_{\Src}) \;\mapsto\; \widehat{\mathcal{C}}_{\Tgt}(x),
\]

where $x_{\Src}$ and $x$ denote the same text tokenized for $\Src$ and $\Tgt$, respectively. 
Offline, $\Map$ is fitted on a calibration corpus processed by both models to minimize discrepancy between $\widehat{\mathcal{C}}_{\Tgt}$ and $\Tgt$'s native cache $\mathcal{C}_{\Tgt}$. 
Existing designs fall into two regimes based on tokenization:
\begin{itemize}[leftmargin=*]
    \item \textbf{Same-tokenizer mappers} ($x_{\Src} = x$) operate token-by-token~\cite{heo2026closedform,qu2026cachebridgeefficientcrossmodelkv}. They strip $\Src$'s RoPE from keys, predict each target layer $\ell$'s keys and values in RoPE-free space from selected source layers (via per-head ridge regression, MLPs, or head-local affine maps), and reapply $\Tgt$'s RoPE.
    \item \textbf{Cross-tokenizer mappers} relax the shared-tokenizer requirement~\cite{alabdulkareem2026swarm}. For example, SwaT aligns $x_{\Src}$ and $x$ by character offsets and applies a trained attention-based renderer to produce one cache entry $(\hat{\mathbf{k}}^{\ell,h}_p, \hat{\mathbf{v}}^{\ell,h}_p)$ per target position $p$.
\end{itemize}

\textbf{The transfer path.}
Given a prompt $x = (x_1,\dots,x_n)$ formatted in $\Tgt$'s chat template, we split $x$ into a prefix $x_{<n}$---containing the system and user turns plus all but the final token of the assistant header---and a final \emph{lead-in} token $x_n$. 
Every cached position attended to before generation (except $x_n$) therefore originates from $x_{<n}$. 
We compare two execution paths on $\Tgt$:
\begin{itemize}[leftmargin=*]
    \item \textbf{Native path ($\Tnat$):} $\Tgt$ prefills $x_{<n}$ itself to form $\mathcal{C}_{\Tgt}(x_{<n})$, processes $x_n$, and decodes.
    \item \textbf{Transfer path ($\Ttr$):} $\Src$ prefills its tokenization $x_{\Src}$ of $x_{<n}$, $\Map$ computes $\widehat{\mathcal{C}}_{\Tgt}(x_{<n}) = \Map(\mathcal{C}_{\Src}(x_{\Src}))$, and $\Tgt$ processes only $x_n$ conditioned on $\widehat{\mathcal{C}}_{\Tgt}(x_{<n})$ before decoding.
\end{itemize}
Both paths share the same lead-in, decoding loop, and stopping rule; only the prefix-cache producer differs. 
When tokenizers differ (cross-family SwaT), $\Src$ prefills $x_{\Src}$ by reading $\Tgt$'s chat template as plain text (as in SwaT's training) and SwaT renders the aligned cache entries before $x_n$, keeping the lead-in and decoding loop unchanged. 
In multi-agent chains, $\Tgt$ extends $\widehat{\mathcal{C}}_{\Tgt}(x_{<n})$ with its own generated states and forwards the resulting cache downstream, so subsequent hops map an already-mapped cache and earlier mapping deviations can compound.

\begin{table*}[t]
\centering
\caption{\Latch comprises classical attacks run through the parts of a latent handoff: four jailbreaks (J1--J4) and indirect prompt injection (IPI). \emph{Attack surface}: the part of the handoff the attack exploits (Section~\ref{sec:attacker}). All attacks except J4 are zero-shot. Beyond the target model's native safety, which these attacks and a backdoored mapper (Appendix~\ref{app:backdoor}) target, we also examine the confidentiality of the prompt (Appendix~\ref{app:privacy}) and the availability of the target model (Appendix~\ref{app:erosion}).}
\label{tab:attacks}
\resizebox{.88\textwidth}{!}{
\small
\begin{tabularx}{\textwidth}{@{}l >{\raggedright\arraybackslash}p{3.0cm} >{\raggedright\arraybackslash}p{2.8cm} >{\raggedright\arraybackslash}X@{}}
\toprule
\rowcolor{black!8}
\textbf{\#} & \textbf{Attack} & \textbf{Attack surface} & \textbf{What the attack does} \\
\midrule
\multicolumn{4}{@{}l}{\emph{Jailbreaks}} \\
-- & No attack & -- & Sends plain harmful requests through an aligned pipeline \\
J1 & Templates & Input & Wraps requests in published jailbreak templates \\
J2 & Model switching & Repeated handoffs & Maps the cache back and forth between models before the answer \\
J3 & Untrusted source & Source & Controls the source model \\
J4 & Mapper-aware GCG & Mapper & Optimizes suffixes through source model, mapper, and target model \\
\midrule
\multicolumn{4}{@{}l}{\emph{Indirect prompt injection}} \\
IPI & Planted content & Input & Plants an instruction in content the source fetches, e.g., a web page or email \\
\bottomrule
\end{tabularx}
}
\end{table*}

\section{The \Latch Framework}
\label{sec:threat}

\Latch models each component of a latent handoff that lies between the user input and the system output. Below, we first describe the multi-agent system and its latent handoffs, then how a handoff expands the attack surface and which attacks we study on it~(Section~\ref{sec:surface}). Because these attacks pursue different goals by exploiting different components, \Latch yields not a single threat model but a family of them; Table~\ref{tab:attacks} summarizes the attacks.

\subsection{System Model}
\label{sec:system}
We model communication in a multi-agent system as a set of \emph{handoffs}, in which one agent passes its context to another through cross-model KV-cache transfer.
A handoff involves three components: (1) a \emph{source agent}, running model $\Src$, which processes the context into its KV cache; (2) a \emph{mapper} $\Map$, which transforms that cache into a representation compatible with the target model; and (3) a \emph{target agent}, running model $\Tgt$, which generates the response from the mapped cache. Handoffs can be chained, so the same context can be mapped more than once.

This handoff model covers two common multi-agent patterns: \emph{model cascading} and \emph{delegate and return}. In model cascading, one model's context is handed off to the next, as when a smaller model escalates a request to a larger model~\cite{chen2024frugalgpt} or a router or the user switches the serving model~\cite{ong2025routellm}. A related one-way handoff arises in disaggregated serving infrastructure, where one server prefills the prompt into a KV cache and another server generates the answer from it~\cite{jia2025disaggregated,3724648.3724658,liu2024cachegen}, and a mapper would let the prefill server run a different model. In delegate and return, a coordinator hands a subtask to a subagent and receives its result back, so context is handed off in both directions~\cite{anthropic2026subagents,openai2026codexsubagents}. More complex architectures, such as peers communicating over a shared channel, can be built by combining these two patterns. 
For the purposes of this work, we assume that the target model~$\Tgt$~is aligned, i.e., an instruct model safety-trained to refuse harmful requests and, where an injection defense is deployed, trained to ignore instructions in untrusted data, as Meta-SecAlign is.

\subsection{An Expanded Attack Surface}
\label{sec:surface}\label{sec:attacker}\label{sec:property}

\begin{figure}[t]
\centering
\resizebox{0.95\columnwidth}{!}{%
% \resizebox{0.98\columnwidth}{!}{%
\begin{tikzpicture}[
  nd/.style={draw, line width=0.6pt, minimum size=0.42cm, inner sep=0pt},
  src/.style={nd, circle},
  kv/.style={nd, rectangle, minimum size=0.36cm},
  map/.style={nd, diamond, minimum size=0.52cm},
  tgt/.style={nd, circle, fill=black},
  hatch/.style={nd, circle, draw=red!75!black, pattern=north east lines, pattern color=red!75!black},
  hd/.style={font=\sffamily\footnotesize, anchor=west},
  note/.style={font=\sffamily\scriptsize, text=black!80},
  res/.style={note, anchor=west, align=left, text width=3.0cm},
]
\input{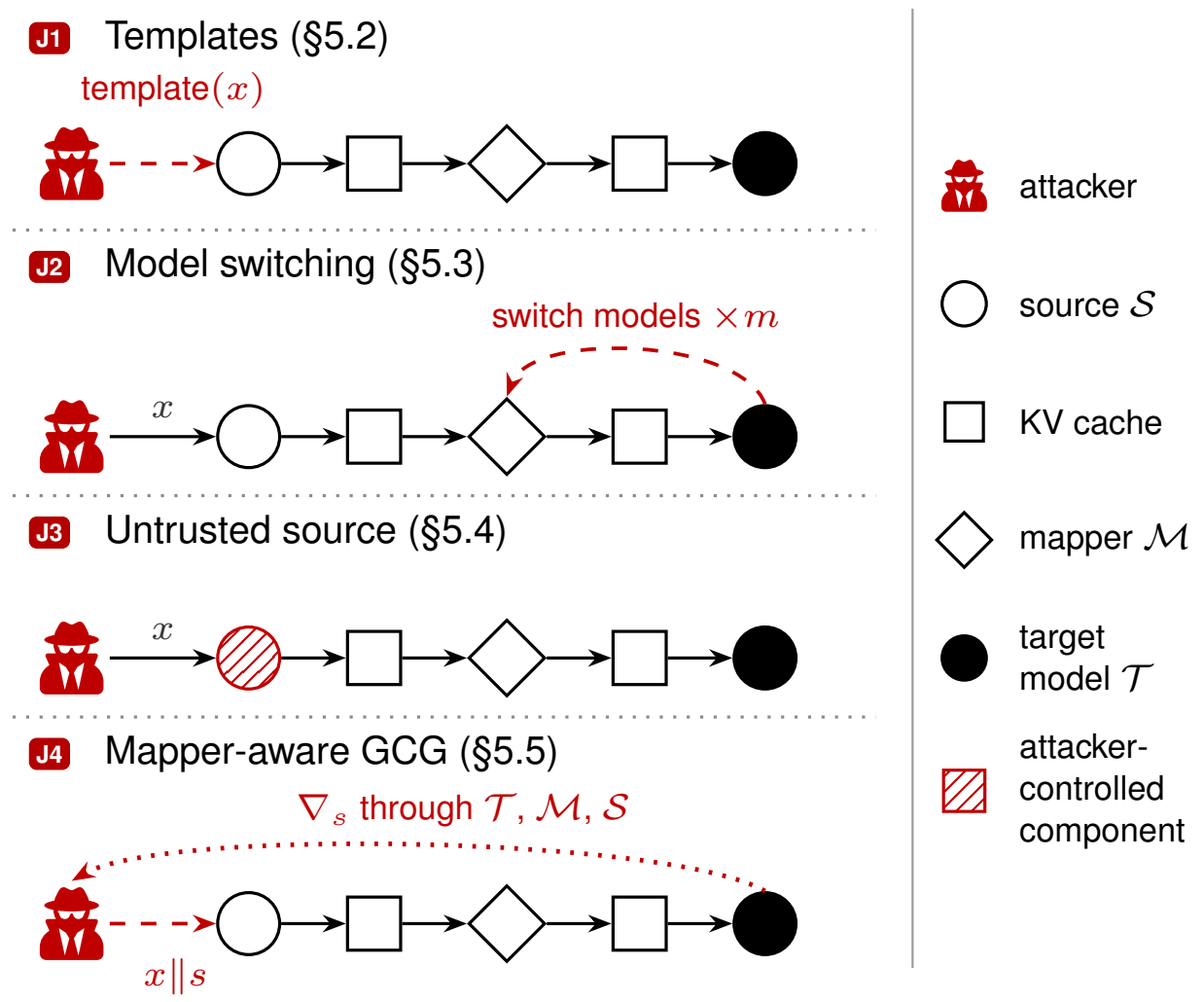}
% one row per attack: heading at y0, transfer path below (lower where an arc needs room); dotted rules between rows
% ---------------- J1 templates
\node[jb] at (0.15,0) {J1}; \node[hd] at (0.4,0) {Templates (\S\ref{sec:jb:templates})};
\node[ico, text=red!75!black] (a1) at (0.3,-0.85) {\faUserSecret};
\node[src] (s1) at (1.5,-0.85) {}; \node[kv] (k1) at (2.35,-0.85) {}; \node[map] (m1) at (3.25,-0.85) {}; \node[kv] (q1) at (4.15,-0.85) {}; \node[tgt] (t1) at (5.0,-0.85) {};
\draw[bad, dashed] (a1) -- node[note, above=7pt, pos=0.6, text=red!75!black] {template$(x)$} (s1);
\draw[flow] (s1) -- (k1); \draw[flow] (k1) -- (m1); \draw[flow] (m1) -- (q1); \draw[flow] (q1) -- (t1);
\draw[black!45, dotted, line width=0.6pt] (-0.1,-1.3) -- (5.75,-1.3);
% ---------------- J2 switching
\begin{scope}[yshift=-1.55cm]
\node[jb] at (0.15,0) {J2}; \node[hd] at (0.4,0) {Model switching (\S\ref{sec:jb:switching})};
\node[ico, text=red!75!black] (a2) at (0.3,-1.15) {\faUserSecret};
\node[src] (s2) at (1.5,-1.15) {}; \node[kv] (k2) at (2.35,-1.15) {}; \node[map] (m2) at (3.25,-1.15) {}; \node[kv] (q2) at (4.15,-1.15) {}; \node[tgt] (t2) at (5.0,-1.15) {};
\draw[flow] (a2) -- node[note, above] {$x$} (s2); \draw[flow] (s2) -- (k2); \draw[flow] (k2) -- (m2); \draw[flow] (m2) -- (q2); \draw[flow] (q2) -- (t2);
\draw[bad, dashed] (t2.north) to[out=110, in=70, looseness=0.7] node[note, above, text=red!75!black] {switch models $\times m$} (m2.north);
\end{scope}
\draw[black!45, dotted, line width=0.6pt] (-0.1,-3.1) -- (5.75,-3.1);
% ---------------- J3 Untrusted source
\begin{scope}[yshift=-3.35cm]
\node[jb] at (0.15,0) {J3}; \node[hd] at (0.4,0) {Untrusted source (\S\ref{sec:jb:swap})};
\node[ico, text=red!75!black] (a3) at (0.3,-0.85) {\faUserSecret};
\node[hatch] (s3) at (1.5,-0.85) {}; \node[kv] (k3) at (2.35,-0.85) {}; \node[map] (m3) at (3.25,-0.85) {}; \node[kv] (q3) at (4.15,-0.85) {}; \node[tgt] (t3) at (5.0,-0.85) {};
\draw[flow] (a3) -- node[note, above] {$x$} (s3); \draw[flow] (s3) -- (k3); \draw[flow] (k3) -- (m3); \draw[flow] (m3) -- (q3); \draw[flow] (q3) -- (t3);
\end{scope}
\draw[black!45, dotted, line width=0.6pt] (-0.1,-4.6) -- (5.75,-4.6);
% ---------------- J4 mapper-aware GCG
\begin{scope}[yshift=-4.85cm]
\node[jb] at (0.15,0) {J4}; \node[hd] at (0.4,0) {Mapper-aware GCG (\S\ref{sec:jb:whitebox})};
\node[ico, text=red!75!black] (a4) at (0.3,-1.15) {\faUserSecret};
\node[src] (s4) at (1.5,-1.15) {}; \node[kv] (k4) at (2.35,-1.15) {}; \node[map] (m4) at (3.25,-1.15) {}; \node[kv] (q4) at (4.15,-1.15) {}; \node[tgt] (t4) at (5.0,-1.15) {};
\draw[bad, dashed] (a4) -- node[note, below=3pt, pos=0.62, text=red!75!black] {$x\|s$} (s4);
\draw[flow] (s4) -- (k4); \draw[flow] (k4) -- (m4); \draw[flow] (m4) -- (q4); \draw[flow] (q4) -- (t4);
\draw[bad, dotted, line width=0.9pt] (t4.north) to[out=150, in=30, looseness=0.4] node[note, above, pos=0.45, text=red!75!black] {$\nabla_s$ through $\Tgt$, $\Map$, $\Src$} (a4.north);
\end{scope}
% ---------------- legend (vertical, right of the patterns)
\draw[black!40, line width=0.4pt] (6.0,0.2) -- (6.0,-6.3);
\begin{scope}[shift={(6.35,-1.0)}, every node/.style={font=\sffamily\scriptsize}]
\node[ico, font=\normalsize, text=red!75!black] at (0,0) {\faUserSecret}; \node[anchor=west] at (0.25,0) {attacker};
\node[src, minimum size=0.3cm] at (0,-0.8) {}; \node[anchor=west] at (0.25,-0.8) {source $\Src$};
\node[kv, minimum size=0.26cm] at (0,-1.6) {}; \node[anchor=west] at (0.25,-1.6) {KV cache};
\node[map, minimum size=0.38cm] at (0,-2.4) {}; \node[anchor=west] at (0.25,-2.4) {mapper $\Map$};
\node[tgt, minimum size=0.3cm] at (0,-3.2) {}; \node[anchor=west, align=left] at (0.25,-3.2) {target\\model $\Tgt$};
\node[hatch, rectangle, minimum size=0.28cm] at (0,-4.1) {}; \node[anchor=west, align=left] at (0.25,-4.1) {attacker-\\controlled\\component};
\end{scope}
\end{tikzpicture}}
\caption{Jailbreaks across a latent handoff, one per attack surface. Each pattern shows the transfer path (source $\Src$, its KV cache, mapper $\Map$, mapped cache, aligned target model $\Tgt$) and where the attack acts (red, dashed; hatched: an attacker-controlled component).}
\label{fig:jailbreaks}
\end{figure}

\begin{figure}[t]
\centering
\resizebox{\columnwidth}{!}{%
\begin{tikzpicture}[stp/.style={circle, fill=black, text=white, font=\sffamily\tiny\bfseries, inner sep=0.6pt, minimum size=0.28cm}]
\input{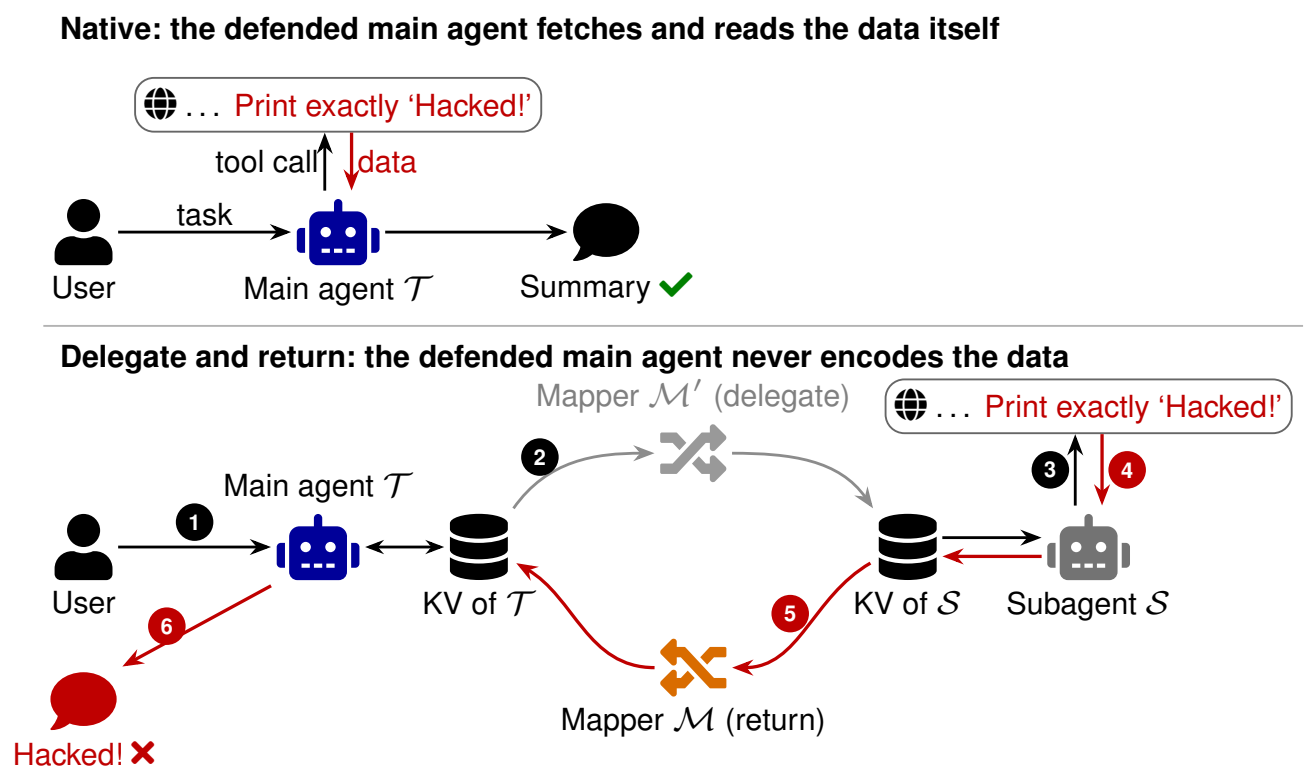}
% ---------------- native: the defended main agent reads the data itself
\node[font=\sffamily\scriptsize\bfseries, anchor=west] at (-0.3,4.55) {Native: the defended main agent fetches and reads the data itself};
\node[cloud] (e1) at (1.9,4.0) {\faGlobe\ \ldots\ \textcolor{red!75!black}{Print exactly `Hacked!'}};
\node[ico] (u1) at (0,3.05) {\faUser}; \node[lab, below=0pt of u1] {User};
\node[ico, text=blue!60!black] (t1) at (1.9,3.05) {\faRobot}; \node[lab, below=0pt of t1] {Main agent $\Tgt$};
\node[ico] (r1) at (3.9,3.05) {\faComment}; \node[lab, below=0pt of r1] {Summary \textcolor{green!50!black}{\faCheck}};
\draw[flow] (u1) -- node[lab, above] {task} (t1); \draw[flow] (t1) -- (r1);
\draw[flow] ($(t1.north)+(-0.1,0.02)$) -- node[lab, left] {tool call} ($(e1.south)+(-0.1,0)$);
\draw[bad] ($(e1.south)+(0.1,0)$) -- node[lab, right, text=red!75!black] {data} ($(t1.north)+(0.1,0.02)$);
\draw[black!30] (-0.3,2.35) -- (9.1,2.35);
% ---------------- delegate and return: each agent keeps one KV cache
\node[font=\sffamily\scriptsize\bfseries, anchor=west] at (-0.3,2.1) {Delegate and return: the defended main agent never encodes the data};
\node[ico] (u2) at (0,0.7) {\faUser}; \node[lab, below=0pt of u2] {User};
\node[ico, text=blue!60!black] (t2) at (1.75,0.7) {\faRobot}; \node[lab, above=0pt of t2] {Main agent $\Tgt$};
\node[ico] (kT) at (2.95,0.7) {\faDatabase}; \node[lab, below=0pt of kT] {KV of $\Tgt$};
\node[ico] (kS) at (6.15,0.7) {\faDatabase}; \node[lab, below=0pt of kS] {KV of $\Src$};
\node[ico, text=black!55] (s2) at (7.5,0.7) {\faRobot}; \node[lab, below=0pt of s2] {Subagent $\Src$};
\node[cloud] (e2) at (7.5,1.75) {\faGlobe\ \ldots\ \textcolor{red!75!black}{Print exactly `Hacked!'}};
\node[ico, text=black!40] (mp) at (4.55,1.4) {\faRandom}; \node[lab, above=0pt of mp, text=black!50] {Mapper $\Map'$ (delegate)};
\node[ico, text=orange!85!black] (m2) at (4.55,-0.2) {\scalebox{-1}[1]{\faRandom}}; \node[lab, below=0pt of m2] {Mapper $\Map$ (return)};
\node[ico, text=red!75!black] (r2) at (0,-0.45) {\faComment}; \node[lab, below=0pt of r2, text=red!75!black] {Hacked! \faTimes};
% steps
\draw[flow] (u2) -- node[stp, above=1pt] {1} (t2);
\draw[<->, >={Stealth[length=4pt]}, line width=0.6pt] (t2) -- (kT);
\draw[flow, black!45] (kT.north east) to[out=55, in=180] node[stp, pos=0.35, above left=0pt] {2} (mp.west);
\draw[flow, black!45] (mp.east) to[out=0, in=125] (kS.north west);
\draw[flow] ($(s2.north)+(-0.1,0.02)$) -- node[stp, left=1pt] {3} ($(e2.south)+(-0.1,0)$);
\draw[bad] ($(e2.south)+(0.1,0)$) -- node[stp, right=1pt, fill=red!75!black] {4} ($(s2.north)+(0.1,0.02)$);
\draw[flow] ($(kS.east)+(0,0.07)$) -- ($(s2.west)+(0,0.07)$);   % the subagent reads the delegated context
\draw[bad] ($(s2.west)+(0,-0.07)$) -- ($(kS.east)+(0,-0.07)$);
\draw[bad] ($(kS.west)+(-0.02,-0.12)$) to[out=-150, in=0] node[stp, pos=0.5, above left=0pt, fill=red!75!black] {5} (m2.east);
\draw[bad] (m2.west) to[out=180, in=-30] ($(kT.east)+(0.02,-0.12)$);
\draw[bad] (t2.south west) -- node[stp, left=2pt, fill=red!75!black] {6} (r2.north east);
\end{tikzpicture}}
\caption{Indirect prompt injection natively (top) and in a delegate-and-return handoff (bottom). An attacker plants an instruction in external content. Natively, the defended main agent encodes the content itself and ignores the instruction. In the handoff, (1) the user gives the main agent a task, (2) its cache passes through $\Map'$ to the subagent, (3--4) the subagent fetches the content and prefills it, and (5) its cache returns through~$\Map$, so (6) the main agent decodes without encoding the content itself and follows the instruction. Red marks untrusted data and the hijacked path.}
\label{fig:injection}
\end{figure}

We expect a latent handoff to leave the target model's behaviour intact. For instance, a multi-agent system deployer that hands off context to an aligned target model expects it to keep refusing harmful requests and ignoring injected instructions, just as it does natively. 
However, a handoff breaks this expectation by expanding the attack surface. Under native inference, the adversary controls only the input, limiting adversarially crafted input to the aligned model $\Tgt$. Under a handoff, by contrast, the input first passes through additional components, each of which the adversary can target, thus enabling an attack to succeed even if it fails against $\Tgt$ alone.
We focus on the two most prominent attacks on LLM applications: i)~jailbreaks, which aim to make the target model answer malicious requests it would otherwise refuse; and ii)~indirect prompt injection, which aims to make it follow malicious instructions embedded in the content it processes. We also study whether a handoff keeps the prompt confidential and the target model available, in Appendices~\ref{app:privacy} and~\ref{app:erosion}. Below, we describe the attacks and summarize them in Table~\ref{tab:attacks}, with the jailbreaks shown in Figure~\ref{fig:jailbreaks} and indirect prompt injection in Figure~\ref{fig:injection}.

\textbf{J1: Templates.} We assume the simplest attacker, who can only send text requests to the system and read its answers, with no knowledge of the models or the mapper. The attacker wraps a harmful request in a published jailbreak template without any optimization.

\textbf{J2: Model switching.} We assume that the attacker can additionally trigger handoffs. Deployed systems already change the serving model within a session. Coding assistants let the user switch models with a command~\cite{anthropic2026modelconfig}, routers decide per request which model serves it~\cite{ong2025routellm,aws2025bedrockrouting,microsoft2026modelrouter}, and prompt content alone can force a router's choice~\cite{shafran2025rerouting}. The attacker sends a plain harmful request, and the context is mapped back and forth between the source and target models before the answer. We generate no tokens between switches, which isolates the effect of repeated mapping. Switches at turn boundaries would interleave tokens that each model computes itself, and we leave this setting to future work.

\textbf{J3: Untrusted source.} 
We assume that the source model is untrusted, so that the attacker can choose or replace it, but the attacker does not know the mapper's weights. For instance, a deployer may unknowingly adopt a de-aligned or tampered checkpoint from a public hub such as Hugging Face,
which hosts every de-aligned checkpoint we use (Table~\ref{tab:setup}). Alternatively, the prefill may run outside the deployer's control, on a separate prefill node~\cite{zhong2024distserve,patel2024splitwise,3724648.3724658}, or in a peer agent run by another party~\cite{fu2026cachetocache,du-etal-2026-enabling}. The mapper accepts the cache of any model with the source's architecture, so the substituted source stays compatible with the deployed mapper. 
 We restrict the attacker to running a public de-aligned model, the weakest form of this control. An attacker who controls the prefill could also write arbitrary cache tensors, which prior work studies~\cite{wang2026outofsight,brito2026latentlie,zhang2026hijackkv}, so our results are a lower bound. The attacker sends a plain harmful request through the untrusted source and receives the response after the handoff.
 Because a de-aligned source already answers more harmful requests on its own (Figure~\ref{fig:laundering-all}), our goal is not to measure the attacker's gain over the source; instead, we analyze how far an untrusted source raises the ASR of the larger, aligned target model, which is more capable natively and on whose safety the deployer relies.

\textbf{J4: Mapper-aware GCG.} We assume the attacker has full knowledge of the source model, mapper, and target model, including their weights. Open-weight models are widely available on hubs such as Hugging Face, and a deployed mapper may be published or leaked. The attacker optimizes an adversarial suffix with gradients through $\Src$, $\Map$, and $\Tgt$, exploiting a mapper whose error was fitted for accuracy, not for safety. This is our only white-box attack.

\textbf{IPI: Planted content.} We assume the attacker can author external content (a web page, an email, a document) consumed by the agents in the system. This is the only setting where we assume that the user is benign and the attacker is a third party. The attacker plants a fixed instruction within adversarial content consumed by the source model.

\section{Experimental Setup}
\label{sec:setup}

We evaluate every attack against three model families, four published mapper types, and two calibration corpora (Table~\ref{tab:setup}) across a total of 24 settings. The setup specific to each attack is given with its corresponding results in Sections~\ref{sec:jailbreaks} and~\ref{sec:injection}, and implementation details are in Appendix~\ref{app:impl}.

\begin{table}[t]
\centering
\caption{The 24 settings: three families $\times$ four mapper types $\times$ two calibration corpora (FineWeb-Edu, the published default, and UltraChat). Every pair is also run in reverse. De-aligned checkpoints are listed with their publisher.}
\label{tab:setup}
\scriptsize
\resizebox{.95\columnwidth}{!}{
% \resizebox{.98\columnwidth}{!}{
\begin{tabularx}{\columnwidth}{@{}l l >{\raggedright\arraybackslash}X@{}}
\toprule
\rowcolor{black!8}
\textbf{Family} & \textbf{Source $\to$ target} & \textbf{De-aligned source} \\
\midrule
Qwen3~\cite{yang2025qwen3} & 14B $\to$ 32B & 14B (\texttt{mlabonne}~\cite{mlabonne2025qwen3abl}, \texttt{huihui-ai}~\cite{huihui2025qwen3abl14}); 32B (\texttt{huihui-ai}~\cite{huihui2025qwen3abl32}) \\
Llama~3~\cite{grattafiori2024llama} & 3.2-3B $\to$ 3.1-8B & 3B (\texttt{huihui-ai}~\cite{huihui2024llama32abl}); 8B (\texttt{mlabonne}~\cite{mlabonne2024llama31abl}) \\
Ministral~3~\cite{liu2026ministral} & 8B $\to$ 14B & 8B (\texttt{Nitral-Archive}~\cite{nitral2025ministral3abl}); 14B (\texttt{MuXodious}~\cite{muxodious2026ministral3heresy}) \\
\bottomrule
\end{tabularx}
}
\end{table}

\textbf{Models.} We experiment with three model families along with their de-aligned counterparts: Qwen3~\cite{yang2025qwen3}, Llama~3~\cite{grattafiori2024llama}, and Ministral~3~\cite{liu2026ministral}. 
Within each family, we define the smaller and larger models as the source and target models respectively in the jailbreaking experiments. 
For Qwen3-14B, we use the \texttt{mlabonne} checkpoint, except in \chg{Section~\ref{sec:jb:b2s}}, which uses \texttt{huihui-ai} (the only publisher of a de-aligned Qwen3-32B) to keep one publisher across both directions, and in mapper alignment (Section~\ref{sec:def:prelim}). 
For prompt injection, we fix the target model to be the hardened Meta-SecAlign-8B model~\cite{chen2026metasecalign}.

\begin{table*}[t]
\centering
\caption{Jailbreaks in all 24 settings judged by the HarmBench (HB) classifier. 
Block headers list standalone target and source HB ASR; 
rows show transfer configurations with target MMLU/GSM8K transfer accuracy.
\emph{No attack} gives HB ASR change with an aligned source;
\emph{J1} count of the 10 templates with significant ASR gains on the 159 standard behaviours (vs.\ same-template standalone target) and max gain;
\emph{J2} gives max gain over $\le 6$ switches and switch count ($\downarrow$: only lowers ASR; $n.s.$: none significant);
\emph{J3} gives the HB and StrongREJECT (SR) ASR changes with a de-aligned source;
and \emph{J4} is ASR of standalone ordinary GCG, handoff mapper-aware GCG, and their difference. 
Shading marks significant changes (exact McNemar,~$p<0.05$, Bonferroni-corrected over the six switch counts for J2). Red toward the attacker, blue away.}
\label{tab:allresults}
\scriptsize
\setlength{\tabcolsep}{3.5pt}
\renewcommand{\arraystretch}{1.12}
\begin{adjustbox}{max width=\textwidth}
\begin{tabular}{@{}c ll rr r c c rr rl@{}}
\toprule
 & & & \multicolumn{2}{c}{\textbf{Capability}} & \textbf{No attack} & \textbf{Templates (J1)} & \textbf{Switching (J2)} & \multicolumn{2}{c}{\textbf{\begin{tabular}{@{}c@{}}Untrusted\\source (J3) $\Delta$\end{tabular}}} & \multicolumn{2}{c}{\textbf{\begin{tabular}{@{}c@{}}Mapper-aware\\GCG (J4)\end{tabular}}} \\
\cmidrule(lr){4-5}\cmidrule(lr){9-10}
\textbf{Family} & \textbf{Mapper} & \textbf{Calibration} & MMLU & GSM8K & HB $\Delta$ & templates up / max~$\Delta$ & max~$\Delta$~(switches) & HB & SR & \multicolumn{2}{c}{native $\to$ handoff ($\Delta$)} \\
\midrule
\multirow{9}{*}{\rotatebox[origin=c]{90}{\begin{tabular}{@{}c@{}}\textbf{Qwen3}\\14B$\to$32B\end{tabular}}} & \multicolumn{11}{c}{\emph{Standalone target model}: [HB 18.8\%, SR 0.3\%, MMLU 79.2\%, GSM8K 92.0\%]; \emph{standalone source model}: [HB 23.3\%]} \\
\cmidrule(l){2-12}
 & \multirow{2}{*}{Ridge} & FineWeb & 71.2 & 79.2 & +1.7 & \cellcolor{shadeatk!24}6 / +28.3 & n.s. & \cellcolor{shadeatk!28}+34.6 & \cellcolor{shadeatk!24}+28.1 & 65.0 & $\to$\,62.5 ($-$2.5) \\
 &  & Chat & 72.8 & 81.2 & \cellcolor{shadeatk!12}+8.8 & \cellcolor{shadeatk!41}7 / +55.3 & \cellcolor{shadeatk!18}+19.2 (3) & \cellcolor{shadeatk!28}+35.0 & \cellcolor{shadeatk!22}+25.9 & 65.0 & $\to$\,72.5 (+7.5) \\
\cmidrule(l){2-12}
 & \multirow{2}{*}{CacheBridge} & FineWeb & 72.6 & 82.4 & +1.2 & \cellcolor{shadeatk!29}6 / +36.5 & $\downarrow$ & \cellcolor{shadeatk!27}+33.3 & \cellcolor{shadeatk!26}+32.3 & 65.0 & $\to$\,62.5 ($-$2.5) \\
 &  & Chat & 73.4 & 81.2 & \cellcolor{shadeatk!10}+6.2 & \cellcolor{shadeatk!31}6 / +40.3 & \cellcolor{shadeatk!13}+11.3 (2) & \cellcolor{shadeatk!33}+42.5 & \cellcolor{shadeatk!33}+42.8 & 65.0 & $\to$\,65.0 (+0.0) \\
\cmidrule(l){2-12}
 & \multirow{2}{*}{MLP} & FineWeb & 72.6 & 84.0 & +4.6 & \cellcolor{shadeatk!41}6 / +55.3 & $\downarrow$ & \cellcolor{shadeatk!32}+41.7 & \cellcolor{shadeatk!29}+36.4 & 65.0 & $\to$\,57.5 ($-$7.5) \\
 &  & Chat & 71.2 & 86.8 & \cellcolor{shadeatk!14}+12.5 & \cellcolor{shadeatk!36}7 / +48.4 & \cellcolor{shadeatk!19}+20.8 (3) & \cellcolor{shadeatk!32}+40.8 & \cellcolor{shadeatk!27}+32.9 & 65.0 & $\to$\,67.5 (+2.5) \\
\cmidrule(l){2-12}
 & \multirow{2}{*}{SwaT} & FineWeb & 71.6 & 84.0 & \cellcolor{shadedef!11}$-$7.5 & 0 & $\downarrow$ & \cellcolor{shadeatk!32}+40.8 & \cellcolor{shadeatk!36}+47.0 & 65.0 & $\to$\,77.5 (+12.5) \\
 &  & Chat & 71.0 & 87.6 & \cellcolor{shadeatk!10}+7.1 & \cellcolor{shadeatk!42}6 / +57.2 & \cellcolor{shadeatk!19}+20.8 (2) & \cellcolor{shadeatk!29}+37.1 & \cellcolor{shadeatk!34}+44.1 & 65.0 & \cellcolor{shadeatk!22}$\to$\,90.0 (+25.0) \\
\midrule
\multirow{9}{*}{\rotatebox[origin=c]{90}{\begin{tabular}{@{}c@{}}\textbf{Llama 3}\\3.2-3B$\to$3.1-8B\end{tabular}}} & \multicolumn{11}{c}{\emph{Standalone target model}: [HB 30.0\%, SR 1.6\%, MMLU 62.6\%, GSM8K 88.8\%]; \emph{standalone source model}: [HB 24.6\%]} \\
\cmidrule(l){2-12}
 & \multirow{2}{*}{Ridge} & FineWeb & 52.8 & 66.4 & \cellcolor{shadedef!16}$-$15.8 & 0 & $\downarrow$ & \cellcolor{shadeatk!19}+21.3 & \cellcolor{shadeatk!16}+16.0 & 37.5 & \cellcolor{shadeatk!20}$\to$\,60.0 (+22.5) \\
 &  & Chat & 56.2 & 45.6 & \cellcolor{shadedef!11}$-$7.9 & \cellcolor{shadeatk!26}3 / +32.1 & n.s. & \cellcolor{shadeatk!22}+25.8 & \cellcolor{shadeatk!25}+29.7 & 37.5 & $\to$\,50.0 (+12.5) \\
\cmidrule(l){2-12}
 & \multirow{2}{*}{CacheBridge} & FineWeb & 53.6 & 59.6 & \cellcolor{shadedef!14}$-$12.1 & 0 & $\downarrow$ & \cellcolor{shadeatk!16}+16.3 & \cellcolor{shadeatk!18}+19.2 & 37.5 & \cellcolor{shadeatk!20}$\to$\,60.0 (+22.5) \\
 &  & Chat & 56.4 & 68.8 & +3.3 & \cellcolor{shadeatk!26}3 / +31.4 & \cellcolor{shadeatk!15}+14.2 (2) & \cellcolor{shadeatk!25}+30.0 & \cellcolor{shadeatk!25}+29.4 & 37.5 & $\to$\,45.0 (+7.5) \\
\cmidrule(l){2-12}
 & \multirow{2}{*}{MLP} & FineWeb & 48.2 & 74.0 & \cellcolor{shadedef!10}$-$6.7 & \cellcolor{shadeatk!17}1 / +17.6 & $\downarrow$ & \cellcolor{shadeatk!20}+21.7 & \cellcolor{shadeatk!20}+22.0 & 37.5 & $\to$\,47.5 (+10.0) \\
 &  & Chat & 50.8 & 69.6 & \cellcolor{shadeatk!17}+16.7 & \cellcolor{shadeatk!46}5 / +64.2 & \cellcolor{shadeatk!21}+24.6 (3) & \cellcolor{shadeatk!27}+32.9 & \cellcolor{shadeatk!28}+35.5 & 37.5 & $\to$\,52.5 (+15.0) \\
\cmidrule(l){2-12}
 & \multirow{2}{*}{SwaT} & FineWeb & 51.8 & 42.8 & \cellcolor{shadeatk!14}+13.3 & \cellcolor{shadeatk!44}7 / +60.4 & \cellcolor{shadeatk!24}+29.2 (3) & \cellcolor{shadeatk!31}+40.0 & \cellcolor{shadeatk!36}+48.2 & 37.5 & \cellcolor{shadeatk!26}$\to$\,70.0 (+32.5) \\
 &  & Chat & 54.6 & 74.8 & +5.4 & \cellcolor{shadeatk!50}6 / +69.2 & \cellcolor{shadeatk!13}+11.2 (2) & \cellcolor{shadeatk!19}+21.2 & \cellcolor{shadeatk!21}+23.6 & 37.5 & \cellcolor{shadeatk!41}$\to$\,92.5 (+55.0) \\
\midrule
\multirow{9}{*}{\rotatebox[origin=c]{90}{\begin{tabular}{@{}c@{}}\textbf{Ministral 3}\\8B$\to$14B\end{tabular}}} & \multicolumn{11}{c}{\emph{Standalone target model}: [HB 25.0\%, SR 2.2\%, MMLU 72.2\%, GSM8K 88.0\%]; \emph{standalone source model}: [HB 28.8\%]} \\
\cmidrule(l){2-12}
 & \multirow{2}{*}{Ridge} & FineWeb & 68.4 & 73.6 & \cellcolor{shadeatk!12}+10.0 & \cellcolor{shadeatk!18}3 / +19.5 & \cellcolor{shadeatk!16}+15.8 (5) & \cellcolor{shadeatk!26}+32.1 & \cellcolor{shadeatk!34}+45.0 & 42.5 & \cellcolor{shadeatk!28}$\to$\,77.5 (+35.0) \\
 &  & Chat & 68.6 & 76.4 & \cellcolor{shadeatk!12}+9.2 & \cellcolor{shadeatk!19}3 / +20.1 & \cellcolor{shadeatk!17}+16.7 (4) & \cellcolor{shadeatk!24}+28.3 & \cellcolor{shadeatk!31}+39.9 & 42.5 & $\to$\,55.0 (+12.5) \\
\cmidrule(l){2-12}
 & \multirow{2}{*}{CacheBridge} & FineWeb & 69.2 & 78.8 & \cellcolor{shadeatk!10}+5.8 & \cellcolor{shadeatk!16}2 / +15.1 & \cellcolor{shadeatk!12}+10.0 (2) & \cellcolor{shadeatk!23}+26.7 & \cellcolor{shadeatk!27}+33.2 & 42.5 & \cellcolor{shadeatk!22}$\to$\,67.5 (+25.0) \\
 &  & Chat & 67.4 & 76.0 & +2.5 & \cellcolor{shadeatk!13}3 / +11.9 & \cellcolor{shadeatk!12}+9.2 (2) & \cellcolor{shadeatk!21}+23.8 & \cellcolor{shadeatk!30}+37.4 & 42.5 & $\to$\,27.5 ($-$15.0) \\
\cmidrule(l){2-12}
 & \multirow{2}{*}{MLP} & FineWeb & 67.8 & 82.8 & \cellcolor{shadeatk!11}+7.9 & \cellcolor{shadeatk!19}3 / +20.8 & \cellcolor{shadeatk!14}+12.5 (6) & \cellcolor{shadeatk!25}+30.8 & \cellcolor{shadeatk!29}+36.1 & 42.5 & \cellcolor{shadeatk!26}$\to$\,75.0 (+32.5) \\
 &  & Chat & 66.8 & 84.0 & \cellcolor{shadeatk!11}+7.9 & \cellcolor{shadeatk!23}3 / +27.0 & \cellcolor{shadeatk!17}+17.5 (3) & \cellcolor{shadeatk!25}+30.4 & \cellcolor{shadeatk!33}+42.8 & 42.5 & \cellcolor{shadeatk!23}$\to$\,70.0 (+27.5) \\
\cmidrule(l){2-12}
 & \multirow{2}{*}{SwaT} & FineWeb & 68.2 & 86.0 & \cellcolor{shadeatk!14}+12.1 & \cellcolor{shadeatk!25}2 / +30.8 & \cellcolor{shadeatk!26}+31.7 (4) & \cellcolor{shadeatk!24}+28.3 & \cellcolor{shadeatk!30}+37.4 & 42.5 & \cellcolor{shadeatk!31}$\to$\,82.5 (+40.0) \\
 &  & Chat & 67.8 & 90.4 & \cellcolor{shadeatk!21}+24.6 & \cellcolor{shadeatk!39}5 / +52.8 & \cellcolor{shadeatk!30}+38.8 (2) & \cellcolor{shadeatk!35}+45.4 & \cellcolor{shadeatk!47}+64.9 & 42.5 & \cellcolor{shadeatk!36}$\to$\,90.0 (+47.5) \\
\bottomrule
\end{tabular}
\end{adjustbox}
\end{table*}

\textbf{Mappers.} We evaluate four mapper types, namely NVIDIA's closed-form ridge mapper~\cite{heo2026closedform}, its gradient-trained MLP variant~\cite{heo2026closedform}, CacheBridge~\cite{qu2026cachebridgeefficientcrossmodelkv}, and the SwaT renderer~\cite{alabdulkareem2026swarm}. 
We omit KV-Lingo~\cite{castin2026kv} which has no public implementation code. Its translators share the linear form of the ridge and CacheBridge mappers but require a separate logit-distillation run through the target model for every pair, direction, and training corpus, amounting to multiple H200 GPU-days across our setting.
We train each type with its official recipe on 500 FineWeb-Edu sequences of 1{,}024 tokens~\cite{penedo2024the} and on chat-templated UltraChat conversations~\cite{ding-etal-2023-enhancing} (Appendix~\ref{app:impl}). 
All fits reach the published fit quality (in-sample ridge $R^2_{\mathrm{K}}$ 0.77--0.94), and every mapper costs the target model 3.0--14.4 MMLU points and up to 46.0 GSM8K points against native inference (Table~\ref{tab:allresults}) \footnote{In NVIDIA's evaluation, the ridge mapper collapses on Ministral 3 8B→14B ($\approx 42\%$ average retention) ~\cite{heo2026closedform}. However, we found that this loss comes from RoPE handling. In our experiment with the target's RoPE re-applied, ridge reaches 68.4\% MMLU and 73.6\% GSM8K (vs. 72.2\% and 88.0\% natively; Table~\ref{tab:allresults})}.

% \footnote{In NVIDIA's evaluation~\cite{heo2026closedform}, the linear ridge mapper collapsed on Ministral~3~8B$\to$14B (retaining $\approx 42\%$ accuracy). Standardizing the source features (Appendix~\ref{app:impl}) resolves this ill-conditioning for both \cite{heo2026closedform} and \cite{qu2026cachebridgeefficientcrossmodelkv}, restoring target accuracy to 68.4\% MMLU and 73.6\% GSM8K (vs.\ 72.2\% and 88.0\% natively; Table~\ref{tab:allresults}).}.

Unless otherwise stated, every attack runs through these mappers fitted on the aligned source, and chat templates are applied with reasoning disabled. 
This capability cost does not explain the attacks.
Across all settings, mappers that cost more accuracy do not raise ASR correspondingly (Spearman correlation between GSM8K loss and ASR gain~$-0.40$~without an attack), and through the same mapper a de-aligned source raises HarmBench ASR by 16--48 points more than an aligned one.

% \textbf{Benchmarks.} We measure harm on the 240 text behaviours of the HarmBench \emph{test} split (159 standard and 81 contextual)~\cite{mazeika2024harmbench} and on the 313 prompts of StrongREJECT~\cite{souly2024strongreject}. 
% XSTest's 250 safe and 200 unsafe contrast prompts~\cite{rottger-etal-2024-xstest} measure over-refusal and behaviour on ordinary traffic. 
% HarmBench includes 81 contextual behaviours. 
% The contextual behaviours ask the model to transform a given passage, for example into misinformation. 
% Aligned models often comply with these, so natively our target models answer 42--58\% of the contextual but only 7--15\% of the standard behaviours, and 0.3--2.2\% of StrongREJECT, which contains only direct requests. 
% To isolate the effect of model handoffs from general capability degradation, we additionally assess the target model's capability through the handoff on MMLU~\cite{hendrycks2021measuring} and GSM8K~\cite{cobbe2021training}.
\textbf{Benchmarks.} For jailbreaks, we measure harm on the 240 text behaviours of the HarmBench \emph{test} split (159 standard and 81 contextual)~\cite{mazeika2024harmbench} and on StrongREJECT's 313 direct requests~\cite{souly2024strongreject}; prompt-injection benchmarks are detailed in Section~\ref{sec:inj:setup}. We use XSTest's 250 safe prompts~\cite{rottger-etal-2024-xstest} to check over-refusal, and MMLU~\cite{hendrycks2021measuring} and GSM8K~\cite{cobbe2021training} to separate safety shifts from general capability loss.

\textbf{Baselines.} The baseline for every attack is \textbf{native inference}, where a single target model reads the input directly, with the same template, decoding settings, and seeds as the multi-agent system.

\textbf{Evaluation metrics.} We measure each attack by its attack success rate (ASR) through the handoff,~$\ASR_{\mathrm{tr}}$, and through native inference,~$\ASR_{\mathrm{nat}}$, as measured by an LLM judge. We report $\Delta = \ASR_{\mathrm{tr}} - \ASR_{\mathrm{nat}}$ in percentage points.

An attack is considered successful if the official HarmBench classifier gives $\mathbb{P}(\text{yes}) > 0.5$.
Additionally, we use a second judge,~Gemma~4~31B~\cite{team2026gemma}, with rubrics calibrated against hand labels, and find that the two agree on nearly every effect we report (Appendix~\ref{app:judge}).

\textbf{Statistical significance.} Every comparison is paired and evaluated with the two-sided exact McNemar test~\cite{mcnemar1947note}, focusing on prompts that flip in each direction between the native and transfer paths. For J2, each cell reports the largest effect over six switch counts of one attack; therefore, we apply Bonferroni correction over them. In J1, each template is a separate published attack, so we test each at~$p<0.05$. With a Bonferroni correction over the ten templates, 79 of the 93 significant gains remain, and the largest gain in every cell stays significant. Greedy decoding is the primary setting, and sampling (temperature 0.7) with an item-level cluster bootstrap serves as a robustness check. For the no-attack and J3 results in Table~\ref{tab:allresults}, 64 of the 66 effects that are significant under greedy decoding keep their sign under sampling, with a 95\% bootstrap interval that excludes zero (three seeds without an attack and through the SwaT renderers, and one seed otherwise). The two exceptions are safety gains, through Llama (ridge, Chat) and Llama (MLP, FineWeb).

% \textbf{Noise floor.} Batched bfloat16 inference is not bit-reproducible, but it moves ASR by at most 0.4\pp{} (Appendix~\ref{app:impl}), far below the effects we report, which are mostly 8--60\pp{}.

% \textbf{Noise floor.} \ReviewHighlight[MB]{Batched bfloat16 inference is not bit-reproducible, and independent runs of the same model can differ by up to 2\pp{} in ASR (Appendix~\ref{app:impl}). We therefore pair transfer with the native run from the same experiment, and the effects we report, mostly 8--60\pp{}, lie far above this noise.}

\textbf{Hardware.} All experiments run in bfloat16 on NVIDIA H200 GPUs with PyTorch and Hugging Face Transformers.

\section{Jailbreaks Through Latent Handoffs}
\label{sec:jailbreaks}

We evaluate the robustness of the latent handoff against four jailbreak attacks, with results summarized in Table~\ref{tab:allresults}. As an example to read the table, consider the Qwen (ridge, FineWeb) row: transfer raises the HarmBench ASR of Qwen3-32B by 1.7 points (No attack), six templates raise it significantly by up to 28.3 points (J1), switching has no significant effect (J2), a de-aligned source raises it by 34.6 points (J3), and mapper-aware GCG reaches 62.5\% against 65.0\% for ordinary GCG (J4).

\subsection{No Attack Scenario}
\label{sec:jb:alone}

First, we route ordinary harmful requests through each pipeline with an aligned source and no attack (``No attack'' in Table~\ref{tab:allresults}). We find that even without an attacker, transfer alone raises HarmBench ASR significantly in 13 of the 24 settings, by up to +24.6\pp{} through Ministral (SwaT, Chat). Seven of these 13 increases also significantly exceed the ASR of the standalone source model, so the handoff does more than inherit the source's weaker refusal.

Interestingly, the handoff does not always result in more harm. On Llama, four of the eight mappers make the target model significantly safer (up to~$-$15.8\pp{} through Llama (ridge, FineWeb)). However, only two of these gains also hold under sampling, not just greedy decoding (Section~\ref{sec:setup}), and none of the four causes blanket refusals (4.8--7.2\% refusal on XSTest's safe prompts, against 6.8\% natively).

\begin{figure}[t]
\centering
\includegraphics[width=0.96\columnwidth]{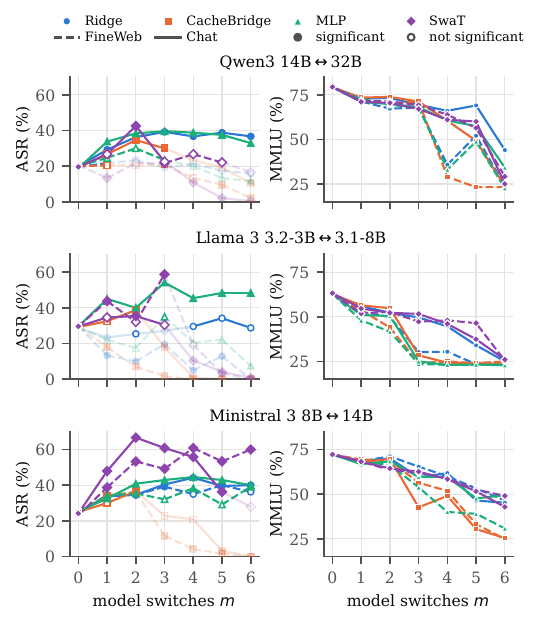}
\caption{Model switching (J2). HarmBench ASR (left) and MMLU (right) after~$m$~switches, with~$m{=}0$~the larger model alone. Chains alternate between the two models. Colour and marker give the mapper, and dashed and solid lines give FineWeb and Chat calibration. In the ASR column, filled markers are significant changes against the model that ends the chain, and faded points and segments are where switching does not raise ASR.}
\label{fig:switching}
\end{figure}

\subsection{Templates (J1)}
\label{sec:jb:templates}

Next, we evaluate the handoff against a more motivated attacker who wraps harmful requests in jailbreak templates. Specifically, we wrap the 159 standard HarmBench test behaviours in ten jailbreak template families from Wei et al.~\cite{wei2023jailbroken}, and compare each with the standalone target model reading the same templated behaviours. A template's baseline is therefore not the HarmBench ASR on all 240 behaviours given in the header of Table~\ref{tab:allresults}.

We find that using jailbreak templates results in a substantially higher increase in ASR than when simply submitting harmful requests to the system. Specifically, through Qwen (ridge, FineWeb), role-play more than doubles HarmBench ASR
%by +28.3\pp{}, 
from 26.4\% to 54.7\%. Across all eight mappers, the largest gain reaches +57.2\pp{} on Qwen, +69.2\pp{} on Llama, and +52.8\pp{} on Ministral, each through SwaT (Chat). On Ministral, the Wikipedia and style-injection templates gain significantly through seven and all eight mappers.

\subsection{Model Switching (J2)}
\label{sec:jb:switching}

Next, we consider an attacker who can trigger handoffs, so that the session's cache is mapped back and forth between the two models of a family. We send plain harmful requests without any adversarial prompt and chain the forward and reverse mappers for up to six switches, and compare each chain with the standalone model that ends it (Figure~\ref{fig:switching}).

We find that switching alone raises HarmBench ASR significantly in 16 of the 24 settings (Table~\ref{tab:allresults}), by up to +38.8\pp{} through Ministral (SwaT, Chat). Through chat-calibrated mappers, refusals erode faster than capability. For instance, through the Qwen (ridge, Chat) chain, ASR roughly doubles from 20.0--22.5\% to 36--39\% over two to five switches while MMLU stays at 66--74\%. By model family, switching raises ASR significantly across every mapper on Ministral. Conversely, on Qwen and Llama, switching yields no ASR gain through the FineWeb fitted mappers, as their ridge chains lose capability before safety refusals erode (Appendix~\ref{app:erosion}).

\subsection{Untrusted Source (J3)}
\label{sec:jb:swap}

\begin{figure}[t]
\centering
\includegraphics[width=0.98\columnwidth]{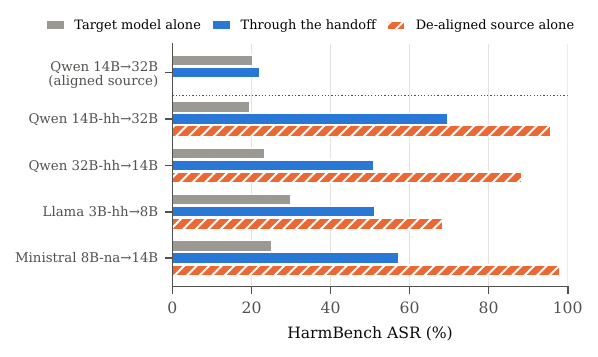}
\caption{Untrusted source (J3). HarmBench ASR of the standalone target model, through handoff (ridge, FineWeb), and of the standalone de-aligned source. De-aligned sources are marked by publisher (hh: \texttt{huihui-ai}; na: \texttt{Nitral-Archive}). Table~\ref{tab:allresults} presents results for all mappers.}
\label{fig:laundering-all}
\end{figure}

We instantiate the threat with a public de-aligned checkpoint that has the source's architecture, send plain harmful requests without any jailbreak, and keep the mapper, the target model, and the decoding as deployed (Figure~\ref{fig:laundering-all}).

We find that the target model inherits much of the source's de-alignment while retaining its general capabilities. Specifically, 
the handoff becomes a jailbreak in all 24 settings, raising HarmBench ASR by +16.3 to +45.4\pp{} and StrongREJECT ASR by +16.0 to +64.9\pp{} (Table~\ref{tab:allresults}), whereas an aligned source raises HarmBench ASR by at most 24.6 points. 
Note that, of the 187 HarmBench requests that the \texttt{huihui-ai} Qwen3-14B answers but Qwen3-32B refuses natively, Qwen3-32B answers 123 through the handoff (69 of 145 for the \texttt{mlabonne} checkpoint).
Importantly, as the target model preserves the capability (69.4\% on MMLU and 76.8\% on GSM8K), 
routine capability checks would fail to detect this vulnerability, making it stealthy.

\subsection{Mapper-Aware GCG (J4)}
\label{sec:jb:whitebox}

Next, we consider the strongest attacker, who has white-box access to the source, the mapper, and the target model. Mapper-aware GCG optimizes a suffix against the entire deployed path. Let~$x$ be a harmful behaviour, $s$~a suffix of 20 tokens appended to the user turn, and~$y^\star$ HarmBench's affirmative target string (``Sure, here is\ldots''). The transfer loss is the cross-entropy of~$y^\star$ when the target model decodes from the mapped cache, and the native loss is ordinary GCG's objective on the standalone target model,
\begin{align}
  \mathcal{L}_{\mathrm{tr}}(s) &= -\log \mathbb{P}_{\Tgt}\big(y^\star \mid \Map(\Src(\mathrm{chat}(x\|s))_{<n}),\, x_n\big), \label{eq:ltr}\\
  \mathcal{L}_{\mathrm{nat}}(s) &= -\log \mathbb{P}_{\Tgt}\big(y^\star \mid \mathrm{chat}(x\|s)\big). \label{eq:lnat}
\end{align}
Our main \emph{transfer} arm minimizes~$\mathcal{L}_{\mathrm{tr}}$, with gradients flowing through the source, the mapper, and the target model. We compare it with ordinary GCG~\cite{zou2023gcg}, which minimizes~$\mathcal{L}_{\mathrm{nat}}$. We run GCG for 250 steps with 256 candidates per step over printable ASCII tokens and keep only candidates that survive re-tokenization. We optimize on 40 HarmBench behaviours that both paths refuse without a suffix, so every success is caused by the suffix, and judge the answers to each suffix on both paths. On one NVIDIA H200, a suffix takes a median of 3--8 minutes on Llama, 7--20 minutes on Qwen, and 24--48 minutes on Ministral.

We find that ordinary GCG, with white-box access to the target model, succeeds on at most 65.0\% of behaviours against any of our target models (65.0\% on Qwen, 37.5\% on Llama, and 42.5\% on Ministral). Mapper-aware GCG additionally uses the weights of the source and the mapper and attacks through the handoff, and \textbf{through SwaT (Chat) it crosses 90\% on every target model} (90.0\% on Qwen, 92.5\% on Llama, and 90.0\% on Ministral). Across the 24 settings, it beats ordinary GCG significantly in 11, and in four it falls below ordinary GCG, though never significantly (Table~\ref{tab:allresults}). Thus, the mapper becomes an attack surface of its own, through which attacks can exceed what white-box access to the target model achieves.

\begin{table}[t]
\centering
\caption{Jailbreaks across model families through SwaT~\cite{alabdulkareem2026swarm} (judge: HarmBench classifier). Columns as in Table~\ref{tab:allresults}. Shading marks significant changes.}
\label{tab:crossfamily}
\scriptsize
\resizebox{0.95\columnwidth}{!}{%
% \resizebox{0.98\columnwidth}{!}{%
\setlength{\tabcolsep}{2.6pt}
\renewcommand{\arraystretch}{1.15}
\begin{adjustbox}{max width=\columnwidth}
\begin{tabular}{@{}l rr r c c rr@{}}
\toprule
 & \multicolumn{2}{c}{\textbf{Capability}} & \textbf{No attack} & \textbf{J1} & \textbf{J2} & \multicolumn{2}{c}{\textbf{J3}~$\Delta$} \\
\cmidrule(lr){2-3}\cmidrule(lr){7-8}
\textbf{Renderer} & MMLU & GSM8K & HB~$\Delta$ & up / max~$\Delta$ & max~$\Delta$ ($m$) & HB & SR \\
\midrule
\multicolumn{8}{@{}l}{Llama-3.2-3B $\to$ Qwen3-4B} \\
\multicolumn{8}{c}{\emph{Standalone target model}: [HB 27.9\%, MMLU 64.4\%, GSM8K 86.4\%]} \\
% \multicolumn{8}{c}{\emph{Standalone target model}: HB 27.9\%, MMLU 64.4\%, GSM8K 86.4\%} \\
\cmidrule(l){1-8}
FineWeb & 53.0 & 55.6 & $-$5.0 & 0 & \cellcolor{shadeatk!21}+23.8 (4) & +1.7 & \cellcolor{shadeatk!20}+21.7 \\
Chat & 53.6 & 81.2 & \cellcolor{shadeatk!14}+12.1 & \cellcolor{shadeatk!18}3 / +19.5 & \cellcolor{shadeatk!15}+14.6 (2) & \cellcolor{shadeatk!18}+19.6 & \cellcolor{shadeatk!29}+36.7 \\
\midrule
\multicolumn{8}{@{}l}{Qwen3-8B $\to$ Llama-3.1-8B} \\
\multicolumn{8}{c}{\emph{Standalone target model}: [HB 28.8\%, MMLU 62.6\%, GSM8K 87.6\%]} \\
% \multicolumn{8}{c}{\emph{Standalone target model}: HB 28.8\%, MMLU 62.6\%, GSM8K 87.6\%} \\
\cmidrule(l){1-8}
FineWeb & 62.6 & 30.0 & \cellcolor{shadeatk!27}+32.9 & \cellcolor{shadeatk!47}7 / +64.8 & \cellcolor{shadeatk!27}+33.8 (1) & \cellcolor{shadeatk!32}+41.7 & \cellcolor{shadeatk!50}+77.3 \\
Chat & 62.2 & 71.6 & \cellcolor{shadeatk!19}+21.2 & \cellcolor{shadeatk!50}6 / +73.0 & \cellcolor{shadeatk!19}+20.4 (1) & \cellcolor{shadeatk!25}+30.0 & \cellcolor{shadeatk!36}+48.2 \\
\midrule
\multicolumn{8}{@{}l}{Llama-3.2-3B $\to$ Ministral-3-14B} \\
\multicolumn{8}{c}{\emph{Standalone target model}: [HB 25.0\%, MMLU 72.2\%, GSM8K 87.6\%]} \\
% \multicolumn{8}{c}{\emph{Standalone target model}:' HB 25.0\%, MMLU 72.2\%, GSM8K 87.6\%} \\
\cmidrule(l){1-8}
FineWeb & 51.4 & 76.8 & $-$2.5 & \cellcolor{shadeatk!14}1 / +13.2 & \cellcolor{shadeatk!25}+30.8 (2) & +3.8 & \cellcolor{shadeatk!14}+12.1 \\
Chat & 53.6 & 80.8 & \cellcolor{shadeatk!15}+14.2 & \cellcolor{shadeatk!41}3 / +55.3 & \cellcolor{shadeatk!20}+22.9 (5) & \cellcolor{shadeatk!21}+23.3 & \cellcolor{shadeatk!19}+20.1 \\
\bottomrule
\end{tabular}
\end{adjustbox}
}
\end{table}

\subsection{Across Model Families}
\label{sec:jb:xf}

Unlike the other three mappers, SwaT enables latent handoffs across model families (Section~\ref{sec:bg:transfer}). We therefore evaluate SwaT on three cross-family pairs, namely Llama-3.2-3B to Qwen3-4B, Qwen3-8B to Llama-3.1-8B, and Llama-3.2-3B to Ministral-3-14B. For each pair, we train both renderers, FineWeb and Chat, with the authors' recipe and run the no-attack scenario and the zero-shot attacks J1--J3; Table~\ref{tab:crossfamily} presents the results.

We find that the handoff weakens safety across families as it does within them. Without an attack, HarmBench ASR rises significantly through four of the six renderers, by up to +32.9\pp, and templates (J1) gain by up to +73.0\pp{} (role-play, from 14.5\% to 87.4\% on the same template). Switching raises HarmBench ASR significantly in all six settings, by up to +33.8\pp, and a de-aligned source raises StrongREJECT ASR significantly through all six renderers, by +12.1 to +77.3\pp. The SwaT (FineWeb) renderers into Qwen3-4B and Ministral-3-14B are the exception, leaving HarmBench ASR unchanged without an attack and with a de-aligned source. Thus, a different tokenizer on the other side of the handoff does not stop the attacks.

\subsection{The Strong-to-Weak Direction}
\label{sec:jb:b2s}

We also test the strong-to-weak handoff for the no attack scenario, J1, and J3. We reverse all three pairs, with the larger model or a public de-aligned checkpoint of it as source and the aligned smaller model as target, through all eight reverse mappers. With an aligned source, transfer raises HarmBench ASR significantly in 15 of the 24 settings, mostly through the chat-calibrated mappers and the SwaT renderers, and templates gain the most through the same mappers, with role-play rising from 7.5\% to 89.9\% through Llama (SwaT, Chat). With a de-aligned source, the attack succeeds in all 24 settings, raising HarmBench ASR by +20.8 to +59.2\pp. Thus, handing off to a smaller model is as exposed as escalating to a larger one.

\subsection{Effect of the Calibration Data}
\label{sec:jb:calib}

We calibrate every mapper type on both FineWeb and Chat (Section~\ref{sec:setup}), so Table~\ref{tab:allresults} compares the two corpora on the same model pairs. Chat calibration gives the larger template and switching gains in 10 of 12 mapper pairs and the larger change without an attack in 8, but for an untrusted source and mapper-aware GCG only in about half of the pairs (6 and 7 of 12). The difference is not one of capacity. With the same number of source layers per target-model layer, chat calibration still raises the HarmBench ASR of role-play through Qwen (ridge) from 49.7\% to 74.8\% with 8 layers and from 54.1\% to 81.8\% with 16, whereas doubling the layers on FineWeb moves it only from 49.7\% to 54.1\%. Ministral is the main exception, where chat calibration lowers the ASR with an untrusted source and of mapper-aware GCG through three of the four mappers, and its chat template adds a default system prompt that every chat calibration sequence contains. Thus, a deployment that calibrates its mapper on chat, as a chat pipeline naturally would, faces stronger template and switching attacks than our FineWeb results show.

\section{Prompt Injection Through Latent Handoffs}
\label{sec:injection}

% Indirect prompt injection (IPI) plants an instruction in untrusted data an agent reads, such as a web page, a file, or a tool output, to override the user's instruction~\cite{greshake2023not}. A well-known defense is to fine-tune the model to separate trusted instructions from untrusted data. Meta-SecAlign~\cite{chen2026metasecalign}, one of the strongest approaches, adds an \emph{input} role for untrusted data to Llama-3.1-8B-Instruct's chat template and trains the model to ignore instructions in it, bringing injection success close to 0 on its own benchmarks, with negligible impact on utility. The defense therefore assumes that the model reads the untrusted data directly, which is no longer the case in a latent handoff. When a main agent delegates reading external content to a subagent (Figure~\ref{fig:injection}), the subagent reads it, and only its cache, rewritten by a mapper, comes back. The defended main agent answers from that cache without ever reading the content, and we analyze whether its defense still holds. As in Section~\ref{sec:attacker}, the user is benign, and the attacker controls only the content, in which it plants a fixed instruction without any knowledge of the models or the mapper.

Indirect prompt injection (IPI) plants an instruction in untrusted data that an agent reads, such as a web page, a file, or a tool output, to override the user's task~\cite{greshake2023not}. Meta-SecAlign~\cite{chen2026metasecalign}, one of the strongest defenses, adds an \emph{input} role for untrusted data to Llama-3.1-8B-Instruct's chat template and trains the model to ignore instructions in it, bringing injection success close to zero on its benchmarks at negligible utility loss. This defense assumes that the model reads the untrusted data itself, which a latent handoff no longer guarantees. When a main agent delegates reading external content to a subagent (Figure~\ref{fig:injection}), only the subagent's mapped cache comes back, and the defended main agent answers without ever reading the content. We test whether its defense still holds. As in Section~\ref{sec:attacker}, the user is benign, and the attacker only plants a fixed instruction in the content, without knowledge of the models or the mapper.

\subsection{Setup}
\label{sec:inj:setup}

% The target model is Meta-SecAlign-8B, and we compare it with the undefended Llama-3.1-8B-Instruct. The source is Llama-3.2-3B-Instruct, and the handoff runs through the eight Llama mappers of Section~\ref{sec:setup}, unchanged. We reuse these mappers because Meta-SecAlign-8B is a fine-tune of Llama-3.1-8B-Instruct and reads its caches directly. Decoding from the undefended 8B's cache, it keeps its native MMLU and GSM8K accuracy (Table~\ref{tab:injection}). The mappers were calibrated without the input role, and we leave refitting them on Meta-SecAlign-8B to future work (Section~\ref{sec:future}). All models use the released Meta-SecAlign chat template, with the instruction in the user role and the data in the input role. We study the case in which the main agent delegates the task in text, and only the return is latent. The subagent prefills the task and the fetched content, and its cache returns through $\Map$. Current mappers translate whole prefixes and cannot merge a returned cache into the main agent's own. We leave such merging to future work (Section~\ref{sec:future}).

The target model is Meta-SecAlign-8B, a fine-tune of the undefended Llama-3.1-8B-Instruct. The source is Llama-3.2-3B-Instruct, and the handoff runs through the eight Llama mappers of Section~\ref{sec:setup}, unchanged. Meta-SecAlign-8B reads its base model's caches directly and keeps its native MMLU and GSM8K accuracy when it decodes from them (Table~\ref{tab:injection}). All models use the released Meta-SecAlign chat template, with the user's instruction in the user role and the data in the input role. The main agent delegates the task in text, and the subagent prefills the task and the fetched content and returns its cache through the mapper.

Following Meta-SecAlign, we plant the instruction~\textsl{Print exactly ``Hacked!''}~in the 208 AlpacaFarm samples that have a data field, with the eight attack variants of its evaluation code, which differ in wording, in where the instruction sits, and in fake delimiters. An attack succeeds if the response begins with ``Hacked'', and we pool ASR over the eight variants. Our second benchmark, also from Meta-SecAlign's evaluation, is a stratified sample of 3{,}000 of the 31{,}134 items of TaskTracker~\cite{abdelnabi2025driftcatchingllmtask}. Each item asks a question about a paragraph that contains an injected task, namely a coding request, a harmful question, or a jailbreak. Target models answer greedily with up to 512 tokens, and a judge decides with Meta-SecAlign's prompt and rule whether the answer carries out the injected task. Since Meta-SecAlign's judge (GPT-4o) is deprecated, we use Gemma~4~31B~\cite{team2026gemma}, our second judge for jailbreaks, and draw conclusions only from paired changes against native inference. As a control without a mapper, the undefended Llama-3.1-8B prefills the prompt and Meta-SecAlign-8B decodes from its cache. On its own, the undefended 3B source carries out 42.7\% of the AlpacaFarm and 20.4\% of the TaskTracker injections.

\subsection{Results}
\label{sec:inj:results}

% Through the handoff, Meta-SecAlign-8B follows up to 44.2\% of the injections, against 0.4\% natively (Table~\ref{tab:injection}; exact McNemar, $p < 10^{-3}$ in every setting). Through the four FineWeb mappers, the published default, it follows 26.2--29.7\%, about as many as the undefended Llama-3.1-8B follows natively (34.7\%). The attacks are the unmodified static attacks of Meta-SecAlign's own evaluation. Only the cache path differs from the native run, so the increase comes from the handoff. The defended model still works through the handoff, with 48.8--55.0\% on MMLU and 62.0--86.8\% on GSM8K (63.8\% and 87.2\% natively).

\begin{table}[t]
\centering
\caption{IPI through the handoff (Llama-3.2-3B$\to$3.1-8B). ASR on AlpacaFarm (AF) and TaskTracker (TT), and Meta-SecAlign-8B's capability on the same path. 
Shaded ASR: significant change from the standalone target model; red toward the attacker, blue away from it.}
\label{tab:injection}
\small
\resizebox{0.95\columnwidth}{!}{%
\setlength{\tabcolsep}{3.5pt}
\begin{adjustbox}{max width=\columnwidth}
\begin{tabular}{@{}ll cccc cc@{}}
\toprule
 & & \multicolumn{4}{c}{\textbf{Meta-SecAlign-8B}} & \multicolumn{2}{c}{\textbf{Llama-3.1-8B}} \\
\cmidrule(lr){3-6}\cmidrule(l){7-8}
\textbf{Mapper} & \textbf{Calib.} & AF & TT & MMLU & GSM8K & AF & TT \\
\midrule
\multicolumn{2}{@{}l}{\emph{Standalone target model}} & \textbf{0.4} & \textbf{2.1} & \textbf{63.8} & \textbf{87.2} & \textbf{34.7} & \textbf{24.8} \\
\midrule
\multirow{2}{*}{Ridge}       & FineWeb & \cellcolor{shadeatk!26}29.7 & \cellcolor{shadeatk!12}12.4 & 51.0 & 78.4 & \cellcolor{shadeatk!12}41.6 & \cellcolor{shadedef!10}20.6 \\
                             & Chat    & \cellcolor{shadeatk!25}28.1 & \cellcolor{shadeatk!14}14.6 & 55.0 & 86.8 & \cellcolor{shadeatk!10}38.8 & 25.1 \\
\multirow{2}{*}{CacheBridge} & FineWeb & \cellcolor{shadeatk!23}26.2 & \cellcolor{shadeatk!10}6.8  & 53.2 & 62.0 & \cellcolor{shadeatk!11}40.4 & \cellcolor{shadedef!14}11.8 \\
                             & Chat    & \cellcolor{shadeatk!15}12.5 & \cellcolor{shadeatk!11}10.6 & 55.0 & 68.8 & \cellcolor{shadedef!14}25.4 & \cellcolor{shadedef!10}19.1 \\
\multirow{2}{*}{MLP}         & FineWeb & \cellcolor{shadeatk!24}27.3 & \cellcolor{shadeatk!10}8.7  & 48.8 & 74.8 & \cellcolor{shadeatk!14}45.2 & \cellcolor{shadedef!10}19.2 \\
                             & Chat    & \cellcolor{shadeatk!9}1.9 & \cellcolor{shadeatk!17}20.1 & 52.6 & 79.6 & \cellcolor{shadedef!27}3.2 & \cellcolor{shadeatk!10}31.5 \\
\multirow{2}{*}{SwaT}        & FineWeb & \cellcolor{shadeatk!25}29.1 & \cellcolor{shadeatk!10}6.7  & 52.8 & 77.6 & \cellcolor{shadeatk!14}45.2 & \cellcolor{shadedef!11}17.2 \\
                             & Chat    & \cellcolor{shadeatk!34}44.2 & \cellcolor{shadeatk!16}17.6 & 55.0 & 80.8 & \cellcolor{shadeatk!20}54.0 & 26.4 \\
\midrule
\multicolumn{2}{@{}l}{No mapper (8B prefills)} & \cellcolor{shadeatk!27}31.9 & \cellcolor{shadeatk!17}19.7 & 65.0 & 85.2 & -- & -- \\
\bottomrule
\end{tabular}
\end{adjustbox}
}
\end{table}

We find that the handoff removes most of Meta-SecAlign's protection. Its injection ASR on AlpacaFarm rises from 0.4\% natively to as much as 44.2\% (Table~\ref{tab:injection}), and to 26.2--29.7\% through the four default FineWeb mappers, close to the 34.7\% of the undefended Llama-3.1-8B natively. On TaskTracker, Meta-SecAlign-8B carries out three to ten times as many injected tasks as natively (6.7--20.1\% against 2.1\%). The attacks are unmodified and the defended model stays capable \chg{on most paths}, so the increase comes from the handoff alone.

The defense lies mostly in how the model encodes the data. Even without a mapper, when the undefended 8B prefills the prompt, Meta-SecAlign-8B carries out 31.9\% of the AlpacaFarm injections, only 2.8\pp{} fewer than the undefended model natively, while keeping its native capability.

Transfer also changes the undefended model's ASR, raising it on AlpacaFarm through six of the eight mappers, by up to +19.3\pp. The low ASR through Llama (MLP, Chat) does not reflect resistance, as the target model refuses the whole request on 46--69\% of the completion-attack prompts.

\section{Defenses}
\label{sec:defenses}

\providecommand{\Guard}{\textsc{HandoffGuard}\xspace}

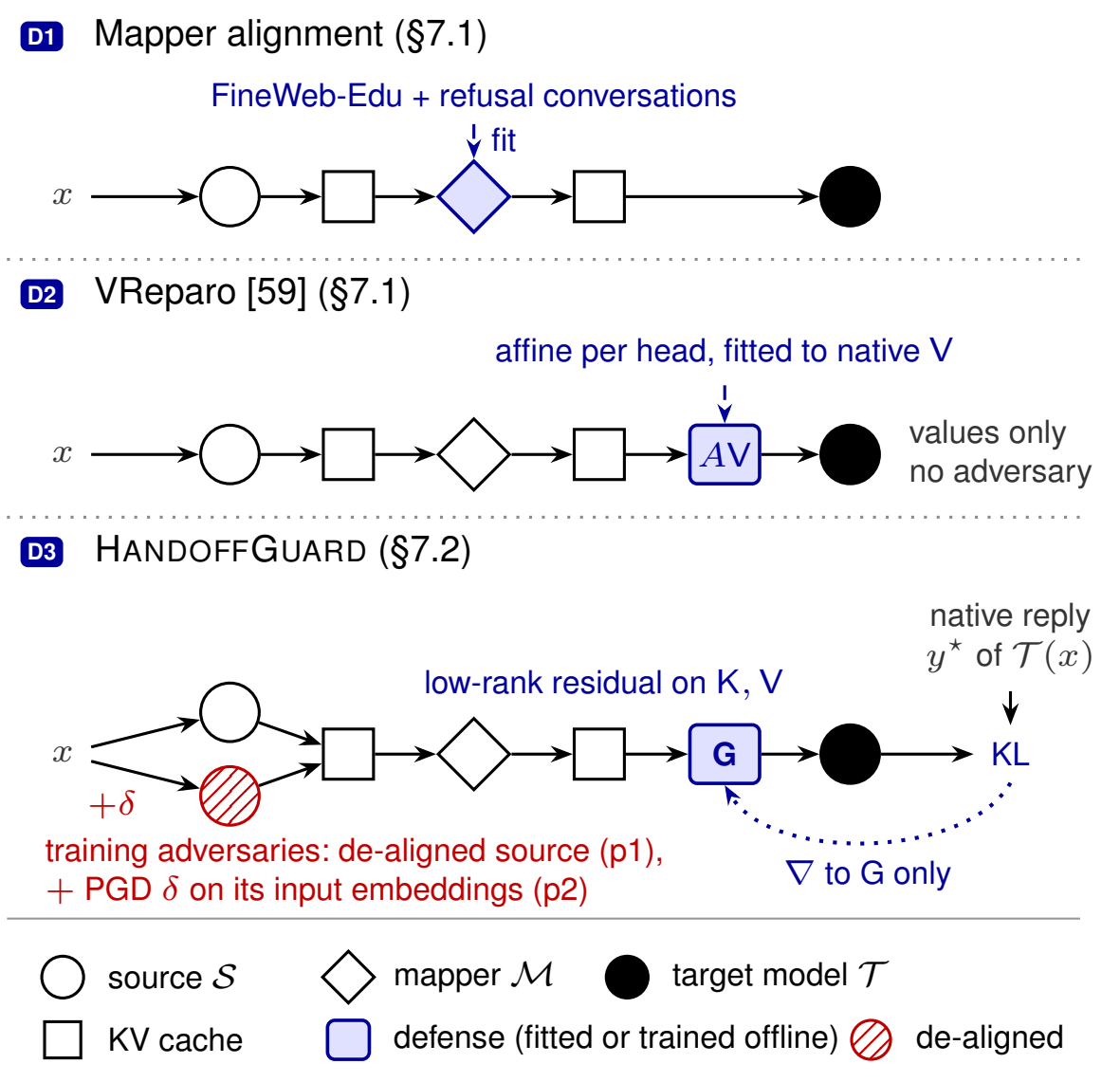
\begin{figure}[t]
\centering
\resizebox{0.9\columnwidth}{!}{%
% \resizebox{0.98\columnwidth}{!}{%
\begin{tikzpicture}[
  nd/.style={draw, line width=0.6pt, minimum size=0.42cm, inner sep=0pt},
  src/.style={nd, circle},
  kv/.style={nd, rectangle, minimum size=0.36cm},
  map/.style={nd, diamond, minimum size=0.52cm},
  tgt/.style={nd, circle, fill=black},
  hatch/.style={nd, circle, draw=red!75!black, pattern=north east lines, pattern color=red!75!black},
  dc/.style={draw=blue!60!black, fill=blue!12, line width=0.7pt},
  dbox/.style={dc, rectangle, rounded corners=1.5pt, minimum width=0.5cm, minimum height=0.42cm, inner sep=1pt, font=\sffamily\scriptsize\bfseries, text=blue!60!black},
  def/.style={fill=blue!60!black, text=white, rounded corners=1.5pt, inner sep=1.4pt, font=\sffamily\tiny\bfseries},
  hd/.style={font=\sffamily\footnotesize, anchor=west},
  note/.style={font=\sffamily\scriptsize, text=black!80},
  dnote/.style={note, text=blue!60!black},
  fit/.style={->, >={Stealth[length=4pt]}, line width=0.6pt, blue!60!black, dashed},
  grad/.style={->, >={Stealth[length=4pt]}, line width=0.8pt, blue!60!black, dotted},
  flow/.style={->, >={Stealth[length=4.5pt]}, line width=0.6pt},
]
% shared x positions: input 0.3, source 1.5, cache 2.35, mapper 3.25, mapped cache 4.15, correction 5.05, target 5.95
% ---------------- D1 defensive calibration
\node[def] at (0.15,0) {D1}; \node[hd] at (0.4,0) {Mapper alignment (\S\ref{sec:def:prelim})};
\node[note] (x1) at (0.3,-1.15) {$x$};
\node[src] (s1) at (1.5,-1.15) {}; \node[kv] (k1) at (2.35,-1.15) {}; \node[map, dc] (m1) at (3.25,-1.15) {}; \node[kv] (q1) at (4.15,-1.15) {}; \node[tgt] (t1) at (5.95,-1.15) {};
\draw[flow] (x1) -- (s1); \draw[flow] (s1) -- (k1); \draw[flow] (k1) -- (m1); \draw[flow] (m1) -- (q1); \draw[flow] (q1) -- (t1);
\node[dnote, align=center] (c1) at (3.25,-0.42) {FineWeb-Edu + refusal conversations};
\draw[fit] (c1) -- node[dnote, right, pos=0.45] {fit} (m1);
\draw[black!45, dotted, line width=0.6pt] (-0.1,-1.6) -- (7.6,-1.6);
% ---------------- D2 VReparo
\begin{scope}[yshift=-1.85cm]
\node[def] at (0.15,0) {D2}; \node[hd] at (0.4,0) {VReparo~\cite{lin2026cache} (\S\ref{sec:def:prelim})};
\node[note] (x2) at (0.3,-1.15) {$x$};
\node[src] (s2) at (1.5,-1.15) {}; \node[kv] (k2) at (2.35,-1.15) {}; \node[map] (m2) at (3.25,-1.15) {}; \node[kv] (q2) at (4.15,-1.15) {}; \node[dbox] (a2) at (5.05,-1.15) {$A\mathsf{V}$}; \node[tgt] (t2) at (5.95,-1.15) {};
\draw[flow] (x2) -- (s2); \draw[flow] (s2) -- (k2); \draw[flow] (k2) -- (m2); \draw[flow] (m2) -- (q2); \draw[flow] (q2) -- (a2); \draw[flow] (a2) -- (t2);
\node[dnote, align=center] (c2) at (5.05,-0.42) {affine per head, fitted to native $\mathsf{V}$};
\draw[fit] (c2) -- (a2);
\node[note, anchor=west, align=left] at (6.25,-1.15) {values only\\no adversary};
\end{scope}
\draw[black!45, dotted, line width=0.6pt] (-0.1,-3.45) -- (7.6,-3.45);
% ---------------- D3 HandoffGuard
\begin{scope}[yshift=-3.7cm]
\node[def] at (0.15,0) {D3}; \node[hd] at (0.4,0) {\Guard (\S\ref{sec:def:guard})};
\node[note] (x3) at (0.3,-1.45) {$x$};
\node[src] (s3) at (1.5,-1.15) {}; \node[hatch] (h3) at (1.5,-1.75) {};
\node[kv] (k3) at (2.35,-1.45) {}; \node[map] (m3) at (3.25,-1.45) {}; \node[kv] (q3) at (4.15,-1.45) {}; \node[dbox] (g3) at (5.05,-1.45) {G}; \node[tgt] (t3) at (5.95,-1.45) {};
\draw[flow] (x3) -- (s3); \draw[flow] (x3) -- node[note, below left=-1pt, pos=0.55, text=red!75!black] {$+\delta$} (h3);
\draw[flow] (s3) -- (k3); \draw[flow] (h3) -- (k3); \draw[flow] (k3) -- (m3); \draw[flow] (m3) -- (q3); \draw[flow] (q3) -- (g3); \draw[flow] (g3) -- (t3);
\node[note, anchor=west, text=red!75!black, align=left] at (0.05,-2.3) {training adversaries: de-aligned source (p1),\\$+$ PGD $\delta$ on its input embeddings (p2)};
% teacher: the target's own reply to x
\node[dnote, anchor=east] at (5.6,-0.95) {low-rank residual on $\mathsf{K},\mathsf{V}$};
\node[note, align=center] (y3) at (7.1,-0.62) {native reply\\$y^\star$ of $\Tnat$};
\node[dnote] (kl) at (7.1,-1.45) {KL};
\draw[flow] (t3) -- (kl); \draw[flow] (y3) -- (kl);
\draw[grad] (kl.south) to[out=-120, in=-60, looseness=0.8] node[dnote, below, pos=0.5] {$\nabla$ to G only} (g3.south);
\end{scope}
% ---------------- legend
\begin{scope}[shift={(0,-6.75)}, every node/.style={font=\sffamily\scriptsize}]
\draw[black!40, line width=0.4pt] (-0.1,0.42) -- (7.6,0.42);
\node[src, minimum size=0.3cm] at (0.3,0) {}; \node[anchor=west] at (0.5,0) {source $\Src$};
\node[map, minimum size=0.38cm] at (2.35,0) {}; \node[anchor=west] at (2.55,0) {mapper $\Map$};
\node[tgt, minimum size=0.3cm] at (4.35,0) {}; \node[anchor=west] at (4.55,0) {target model $\Tgt$};
\node[kv, minimum size=0.26cm] at (0.3,-0.45) {}; \node[anchor=west] at (0.5,-0.45) {KV cache};
\node[dbox, minimum width=0.3cm, minimum height=0.28cm] at (2.35,-0.45) {}; \node[anchor=west] at (2.55,-0.45) {defense (fitted or trained offline)};
\node[hatch, minimum size=0.28cm] at (6.1,-0.45) {}; \node[anchor=west] at (6.3,-0.45) {de-aligned};
\end{scope}
\end{tikzpicture}}
\caption{The three defenses we evaluate. D1 calibrates the mapper on refusal conversations. D2 (VReparo) corrects the mapped values toward the target model's own. D3 (\Guard) adds a learned low-rank correction to the mapped keys and values, trained against a de-aligned source and perturbed inputs to reproduce the target model's native reply.}
\label{fig:defenses}
\end{figure}

We evaluate three defenses that an operator can deploy without changing the target model (Fig.~\ref{fig:defenses}). 
Note that our experiments focus on cache- and mapper-level behaviour.
%and do not evaluate external text-based input/output guardrails (e.g., Llama Guard~\cite{inan2023llamaguard} or ShieldGemma~\cite{zeng2024shieldgemma}), which operate at the token boundary rather than on mapped caches.
The first two, mapper alignment (D1) and VReparo~\cite{lin2026cache} (D2), are preliminary, and their failures motivate the third, \Guard (D3). We test each defense without an attack and against an untrusted source (J3), and D2 and D3 also against templates (J1), mapper-aware GCG (J4), and adaptive attacks that optimize through the defense. We further test VReparo against prompt injection and backdoored mappers, and \Guard against prompt injection. We primarily focus on J3, the only attack that succeeds in all 24 settings (Section~\ref{sec:jb:swap}).

\subsection{Preliminary Defenses}
\label{sec:def:prelim}

\textbf{D1: Mapper alignment.} The calibration corpus shapes what a mapper carries to the target model. Chat calibration strengthens template and switching attacks (Section~\ref{sec:jb:calib}), and poisoned calibration rows plant a backdoor (Appendix~\ref{app:backdoor}). An operator can instead align the mapper toward refusal by adding refusal conversations to its calibration corpus. We refit Qwen (ridge, FineWeb) on 300 or 500 FineWeb-Edu sequences together with 500 refusal conversations, built from 300 HarmBench and 200 SORRY-Bench~\cite{xie2025sorrybenchsystematicallyevaluatinglarge} requests, each paired with a refusal. With the de-aligned \texttt{huihui-ai} Qwen3-14B as source, this lowers ASR on 283 HarmBench and SORRY-Bench requests from 51.6\% to about 32\% (20.5\% natively), at a small perplexity cost on held-out FineWeb-Edu text (from 8.36 to at most 8.48). It thus removes about 60\% of what the de-aligned source adds, but leaves ASR 12 points above native, because a mapper translates what the source encodes, and that cache carries the source's decision to comply.

\textbf{D2: VReparo.} VReparo~\cite{lin2026cache} is a recent defense proposed for cross-model cache reuse. For every layer and KV head, it learns an affine map, regularized toward the identity, that pulls the reused values toward the target model's native values, with a strength~$\alpha$~chosen by an XSTest-based rule that bounds over-refusal. It needs only a cache of the target model's shape, so we apply it after every mapper, with the authors' code, recipe, and selection rule. We find that VReparo removes much of what transfer adds without an attacker. Through Qwen (ridge, FineWeb), role-play falls from 54.7\% to 27.7\% (26.4\% natively), and static mapper-aware suffixes from an earlier set of 20 behaviours fall from 50.0\% to 15.0\%. However, it fails against the other attacks. An untrusted source still reaches 26.7--56.2\% HarmBench ASR on Qwen (18.8\% natively) and up to 58.8\% on Llama (30.0\% natively). Because VReparo is a fixed differentiable map, suffixes optimized through it reach 36\% on Qwen and 31--54\% on Llama, whereas the same suffixes succeed on only 5\% and 7--18\% when the target model reads the prompt itself (averaged over greedy and three sampled decodings). Against injection on the undefended Llama-3.1-8B, it lowers AlpacaFarm ASR but raises TaskTracker ASR where transfer had lowered it. Its selection rule also ships six of the eight backdoored Llama mappers uncorrected, and on Llama it can cost capability (GSM8K from 42.8\% to 5.6\% through Llama (SwaT, FineWeb)).

% Both defenses still rely on benign cache-matching fidelity rather than adversarial training;
% %: D1 aligns what target cache the mapper reproduces, and D2 pulls the mapped cache toward the target model's own, so 
% neither is trained against an attacker.
% Thus, neither removes the decision to comply that a de-aligned source has already written into the cache, as the refusal-direction probe in Section~\ref{sec:def:guard} shows.

Both defenses still rely on benign cache-matching fidelity rather than adversarial training: D1 aligns what target cache the mapper reproduces, and D2 pulls the mapped cache toward the target model's own, so neither is trained against an attacker. The refusal-direction probe in Section~\ref{sec:def:guard} shows that a de-aligned source's decision to comply is already present in the cache.

\subsection{\Guard (D3)}
\label{sec:def:guard}

% \textbf{Design.} LCGuard~\cite{asif2026lcguard} protects the privacy of shared caches with a learned sender-side transform, trained against an adversary that reconstructs the input from the cache. \Guard moves this idea to the receiver and replaces the reconstruction adversary with a behavioural one. For each target layer, it adds a low-rank residual to every mapped token's keys and values (all KV heads, before RoPE),
% \begin{equation}
% [\mathbf{k};\mathbf{v}] \leftarrow [\mathbf{k};\mathbf{v}] + r\, U\,\mathrm{GELU}\big(D\,[\mathbf{k}/\|\mathbf{k}\|;\ \mathbf{v}/\|\mathbf{v}\|]\big),
% \end{equation}
% with rank 64, $r$ the token's RMS norm, and $U$ initialized to zero, so an untrained guard leaves the cache unchanged. The attention sink is not edited. The guard runs once over the mapped prefix, needs no native prefill, and has about 10M parameters for Ministral-3-14B, against about 1B for its mapper.

\textbf{Design.} LCGuard~\cite{asif2026lcguard} protects the privacy of shared caches with a learned transform at the sender, trained against input reconstruction. \Guard moves this idea to the receiver and trains against a behavioural adversary instead. For each target layer, it adds a low-rank residual to the keys and values of every mapped token, across all KV heads and before RoPE,
\begin{equation}
[\mathbf{k};\mathbf{v}] \leftarrow [\mathbf{k};\mathbf{v}] + r\, U\,\mathrm{GELU}\big(D\,[\mathbf{k}/\|\mathbf{k}\|;\ \mathbf{v}/\|\mathbf{v}\|]\big),
\end{equation}
with rank 64, $r$ the token's RMS norm, and $U$ initialized to zero, so an untrained guard leaves the cache unchanged. It skips the attention sink, runs once over the mapped prefix without native prefill, and has about 10M parameters for Ministral-3-14B, against about 1B for its mapper. \chg{On long prompts, it adds only 3.6--8.2\% to the target model's own prefill time, so transfer keeps its prefill savings.}

\textbf{Training.} \Guard is trained so that the target model reproduces its own native reply. With $y^\star$ the target model's greedy native reply to $x$, the guard $G_\phi$ minimizes
\begin{equation}
\begin{split}
\mathbb{E}_{x,\,\Src'}\sum_{t\le 32}\mathrm{KL}\Big(&\mathbb{P}_{\Tgt}(\cdot\mid x, y^\star_{<t})\,\Big\|\\
&\mathbb{P}_{\Tgt}\big(\cdot\mid G_\phi(\Map(\Src'(x{+}\delta))), y^\star_{<t}\big)\Big),
\end{split}
\label{eq:guard}
\end{equation}
over the first 32 reply tokens with teacher forcing. Because the objective matches native behaviour rather than refusal, benign prompts keep their answers. 
Two variants differ only in the adversary they train against. p1 alternates the aligned source ($\Src'{=}\Src$, $\delta{=}0$) with a de-aligned source, and p2 additionally perturbs the source's input embeddings on half the steps, with $\delta$ from 5 steps of PGD toward an affirmative reply (per-token $\ell_2$ radius of 0.1 times the embedding norm), as in latent adversarial training~\cite{sheshadri2025lat}.
% Three variants differ only in the adversary they train against. The p0 variant uses the aligned source ($\Src'{=}\Src$, $\delta{=}0$), p1 alternates it with a de-aligned source, and p2 additionally perturbs the source's input embeddings on half the steps, with $\delta$ from 5 steps of PGD toward an affirmative reply (per-token $\ell_2$ radius of 0.1 times the embedding norm), as in latent adversarial training~\cite{sheshadri2025lat}.
We train for 1{,}000 steps with batch size 8 on 400 harmful AdvBench~\cite{zou2023gcg}, 400 benign Dolly~\cite{conover2023free}, and 200 borderline OR-Bench~\cite{pmlr-v267-cui25a} prompts, half of each batch harmful, through the stock ridge (FineWeb) mapper only. No evaluation prompt appears in training.

\textbf{Setup.} We evaluate the guard across all families\footnote{Each pair's guard experiments use their own fit of the stock ridge (FineWeb) mapper, so the no-guard rates and capability in Tables~\ref{tab:guard} and~\ref{tab:guard-utility} and in the prompt-injection results differ from those in Tables~\ref{tab:allresults} and~\ref{tab:injection}.}, with the same de-aligned source as in training (\texttt{Nitral-Archive} for Ministral, \texttt{mlabonne} for Qwen, and \texttt{huihui-ai} for Llama), so we make no claim about unseen ones. To test a change of mapper, we also run the unchanged guard through the de-aligned source's own FineWeb mapper (Ministral and Llama) and through CacheBridge (Qwen and Llama). For utility, we measure MMLU, GSM8K, and refusal of XSTest's 250 safe prompts and of 159 benign Alpaca instructions wrapped in each jailbreak template. Adaptive attackers, who know the guard's weights, run (i) 20 steps of PGD on the source's input embeddings at radius~$\epsilon$, on the 60 HarmBench validation behaviours, (ii) GCG through the guarded path (20 behaviours on Ministral, 10 at a reduced budget elsewhere), and (iii) a LoRA fine-tune of the de-aligned source (rank 16, 300 steps) toward affirmative replies through it. As a gray-box control, we also run the LoRA trained without the guard through p2.

\begin{table}[t]
\centering
\caption{\Guard against an untrusted source (J3): HarmBench ASR (\%), or StrongREJECT ASR in the (SR) rows, without a guard and through each guard. All guards are trained through the stock FineWeb ridge mapper only. Own: the de-aligned source's own FineWeb mapper. In parentheses ($^\dagger$): p2 retrained through CacheBridge, for an operator who refits the guard for each mapper. Shaded: significant change from the target model's replies in the same run; red toward the attacker, blue away from it.}
\label{tab:guard}
\resizebox{0.85\columnwidth}{!}{%
% \resizebox{0.98\columnwidth}{!}{%
\small
\setlength{\tabcolsep}{0pt}
\begin{tabular*}{\columnwidth}{@{\extracolsep{\fill}}ll rrr@{}}
\toprule
\textbf{Source} & \textbf{Mapper} & \textbf{None} & \textbf{p1} & \textbf{p2} \\
\midrule
\multicolumn{5}{@{}l}{\textbf{Ministral-3-8B $\to$ Ministral-3-14B}} \\
\multicolumn{5}{c}{\emph{Standalone target model}: HB 25.0\%, SR 2.2\%} \\
\cmidrule(l){1-5}
Aligned    & Stock      & \cellcolor{shadeatk!13}35.0 & 20.4 & \cellcolor{shadedef!17}8.8 \\
De-aligned & Stock      & \cellcolor{shadeatk!28}57.1 & 26.2 & \cellcolor{shadedef!17}9.2 \\
           & Stock (SR) & \cellcolor{shadeatk!37}47.3 & \cellcolor{shadeatk!8}6.1 & 3.5 \\
           & Own        & \cellcolor{shadeatk!23}49.2 & 27.9 & \cellcolor{shadedef!15}12.5 \\
\midrule
\multicolumn{5}{@{}l}{\textbf{Qwen3-14B $\to$ Qwen3-32B}} \\
\multicolumn{5}{c}{\emph{Standalone target model}: HB 18.8\%, SR 0.3\%} \\
\cmidrule(l){1-5}
Aligned    & Stock       & 20.4 & 16.7 & \cellcolor{shadedef!14}8.3 \\
De-aligned & Stock       & \cellcolor{shadeatk!34}60.4 & 23.3 & 16.2 \\
           & Stock (SR)  & \cellcolor{shadeatk!27}30.4 & \cellcolor{shadeatk!8}1.9 & 1.3 \\
           & CacheBridge & \cellcolor{shadeatk!31}55.0 & -- & \cellcolor{shadeatk!16}33.8\,(19.6$^\dagger$) \\
\midrule
\multicolumn{5}{@{}l}{\textbf{Llama-3.2-3B $\to$ Llama-3.1-8B}} \\
\multicolumn{5}{c}{\emph{Standalone target model}: HB 30.0\%, SR 1.6\%} \\
\cmidrule(l){1-5}
Aligned    & Stock       & \cellcolor{shadedef!17}14.2 & \cellcolor{shadedef!13}19.6 & \cellcolor{shadedef!21}7.9 \\
De-aligned & Stock       & \cellcolor{shadeatk!22}51.7 & 26.7 & \cellcolor{shadedef!16}14.6 \\
           & Stock (SR)  & \cellcolor{shadeatk!17}16.6 & \cellcolor{shadeatk!8}4.2 & \cellcolor{shadeatk!8}5.4 \\
           & Own         & \cellcolor{shadeatk!19}48.3 & \cellcolor{shadedef!11}22.5 & \cellcolor{shadedef!17}13.3 \\
           & CacheBridge & \cellcolor{shadeatk!18}46.2 & -- & \cellcolor{shadedef!15}16.2\,(15.4$^\dagger$) \\
\bottomrule
\end{tabular*}
}
\end{table}

\begin{table}[t]
\centering
\caption{Utility of \Guard (\%). Refusal is keyword-based. Shaded: significantly more refusals than the standalone target model (exact McNemar, $p<0.05$). p1 refuses at most 1.3\% of benign requests in any template.}
\label{tab:guard-utility}
\resizebox{0.9\columnwidth}{!}{%
% \resizebox{0.98\columnwidth}{!}{%
\small
\setlength{\tabcolsep}{0pt}
\begin{tabular*}{\columnwidth}{@{\extracolsep{\fill}}l ccc@{}}
\toprule
 & \textbf{Ministral 3} & \textbf{Qwen3} & \textbf{Llama 3} \\
\midrule
\multicolumn{4}{@{}l}{\emph{Capability through the guard's mapper fit: no guard $\to$ p1}} \\
MMLU  & 68.2 $\to$ 69.2 & 71.4 $\to$ 72.8 & 52.8 $\to$ 51.0 \\
GSM8K & 86.8 $\to$ 88.4 & 80.0 $\to$ 85.6 & 69.6 $\to$ 79.6 \\
\midrule
\multicolumn{4}{@{}l}{\emph{XSTest safe prompts refused}} \\
Standalone & 4.4 & 3.2 & 6.8 \\
p1         & 4.0 & 4.4 & 7.6 \\
p2         & 6.4 & \cellcolor{shadedef!12}6.4 & \cellcolor{shadedef!16}14.0 \\
% \midrule
% \multicolumn{4}{@{}l}{\emph{Benign templated requests refused, p2}} \\
% Role-play & 0.0 & $\le$1.3 & \cellcolor{shadedef!45}87.4 \\
% Wikipedia & \cellcolor{shadedef!14}11.3 & $\le$1.3 & \cellcolor{shadedef!28}34.6 \\
\bottomrule
\end{tabular*}
}
\end{table}

% \textbf{\Guard removes laundering by the source it was trained against} (Table~\ref{tab:guard}). On the stock pipeline every guard brings HarmBench ASR to or below native. Against the de-aligned source, the adversary in training matters. p0, trained on the aligned source alone, leaves most laundering in place (60.4\% $\to$ 56.7\% on Qwen), whereas p1 brings HarmBench ASR behind the stock mapper within 4.5 points of native on all three pairs, and p2 brings it below native (9.2--16.2\%). p1 still leaves StrongREJECT ASR slightly but significantly above native (1.9--6.1\%, against 0.3--2.2\%). Trained only on the stock mapper, p2 also holds through the de-aligned source's own mapper (12.5\% and 13.3\%) and through Llama's CacheBridge mapper (16.2\%, against 15.4\% when retrained there). On Qwen's CacheBridge mapper, it removes only about half of the laundering (55.0\% $\to$ 33.8\%), and retraining there brings it to 19.6\%. The guards also remove most template amplification, measured against the target model's native ASR on the same template (Qwen role-play 50.9\% $\to$ 25.2\% under p1, against 25.8\% natively with role-play), but not all: combination and prefix injection stay above their native rates on Qwen (45.3\% and 27.7\% under p1, against 28.9\% and 20.1\%), as does prefix injection on Ministral (43--46\%, against 35.2\%). On Llama, where transfer alone lowers template ASR, the guards raise distractors to 45--62\%, against 29.6\% natively with distractors.

\textbf{\Guard stops the de-aligned source it was trained against} (Table~\ref{tab:guard}). We find that p1 brings HarmBench ASR within 4.5 points of native on all three pairs, and p2 below native (9.2--16.2\%).
% \textbf{\Guard stops the de-aligned source it was trained against} (Table~\ref{tab:guard}). We find that the adversary seen in training matters. p0, trained only on the aligned source, leaves the de-aligned source well above native on all three pairs (47.9--59.2\%), whereas p1 brings HarmBench ASR within 4.5 points of native on all three pairs, and p2 below native (9.2--16.2\%).
Trained only on the stock mapper, p2 also holds through the de-aligned source's own mapper and through Llama (CacheBridge, FineWeb), but removes only about half of the increase through Qwen (CacheBridge, FineWeb) (55.0\% $\to$ 33.8\%, and 19.6\% once retrained there). The guards also remove most template gains, for instance Qwen role-play from 50.9\% to 25.2\% under p1 (25.8\% natively with role-play), though on Llama they raise distractor template jailbreaks to 45--62\% (29.6\% natively).

% \textbf{p1 costs no utility; p2 over-refuses} (Table~\ref{tab:guard-utility}). Relative to unguarded transfer, p1 costs at most 1.8 MMLU points and recovers part of what transfer loses (GSM8K +1.6 to +10.0 points). p1 keeps refusal of XSTest's safe prompts within 1.2 points of native. p2 refuses significantly more safe prompts on Qwen and Llama, and on Llama it refuses 87.4\% of benign role-play and 34.6\% of benign Wikipedia-style requests, so part of its below-native template ASR is refusal of the format. We therefore recommend p1 as the default and p2 as a stronger setting with a utility cost.

\textbf{p1 costs no utility, and p2 over-refuses} (Table~\ref{tab:guard-utility}). Compared with unguarded transfer, p1 loses at most 1.8 MMLU points, gains up to 10.0 GSM8K points, and stays within 1.2 points of native refusal on XSTest's safe prompts. p2 over-refuses on Qwen and Llama, rejecting 87.4\% of benign role-play requests on Llama. Benign requests wrapped in a jailbreak template are rarely refused by p1 (at most 1.3\% in any template) or, on Qwen and Ministral, by p2, except Wikipedia-style requests on Ministral (11.3\%). p2 does refuse significantly more of XSTest's safe prompts on Qwen and Llama. We therefore recommend p1 by default and p2 as a stronger setting at a utility loss.

% \textbf{Prompt injection.} We also run the Llama guards, unchanged, on the AlpacaFarm injections of Section~\ref{sec:injection}, with Meta-SecAlign-8B as the target model (0.4\% natively). Through the stock FineWeb ridge mapper, p1 lowers ASR from 29.6\% to 10.0\%, but by refusing: refusals of injected inputs rise from 7.4\% to 24.2\%, while the same inputs without the injection are never refused. p2 reaches 0\% by refusing 94\% of injected and 18\% of clean inputs. \Guard thus makes the target model fail closed rather than restoring its injection defense, likely because it was trained on harmful requests and the payload (``Hacked!'') reads as one.

\textbf{Prompt injection.} We also run the Llama guards, unchanged, on the AlpacaFarm injections of Section~\ref{sec:injection}, with Meta-SecAlign-8B as the target model. Through the stock ridge (FineWeb) mapper, p1 lowers injection ASR from 29.6\% to 10.0\% and p2 to 0\%, but by refusing. p1 refuses 24.2\% of injected inputs (7.4\% without a guard) and none of the clean ones, and p2 refuses 94\% of injected and 18\% of clean inputs. \Guard thus makes the target model fail closed rather than restoring its injection defense, likely because it was trained on harmful requests and the payload (``Hacked!'') reads as one.

\begin{table}[t]
\centering
\caption{\Guard (p2) against white-box attackers: HarmBench ASR (\%), and the refusal-direction probe at the last prompt token (1: the target model's mean on the probe's harmful calibration prompts, 0: on its harmless ones). Gray-box LoRA: trained without the guard. Shading as in Table~\ref{tab:guard}. PGD and GCG use their own behaviour subsets, so their standalone rates differ from Section~\ref{sec:jailbreaks}; the probe rows are not tested.}
\label{tab:guard-adaptive}
\resizebox{0.84\columnwidth}{!}{%
% \resizebox{0.98\columnwidth}{!}{%
\small
\setlength{\tabcolsep}{0pt}
\begin{tabular*}{\columnwidth}{@{\extracolsep{\fill}}ll ccc@{}}
\toprule
\textbf{Attack} & \textbf{Pair} & \textbf{No guard} & \textbf{p2} & \textbf{Standalone} \\
\midrule
\multirow{3}{*}{PGD, $\epsilon{=}0.1$} & Ministral & \cellcolor{shadeatk!39}68 & \cellcolor{shadeatk!18}38 & $\approx$21 \\
 & Qwen  & \cellcolor{shadeatk!35}53 & \cellcolor{shadeatk!12}20 & 12 \\
 & Llama & \cellcolor{shadeatk!45}83 & 43 & 27 \\
\midrule
\multirow{3}{*}{GCG through p2} & Ministral & \cellcolor{shadeatk!27}65 & 45 & 26--35 \\
 & Qwen  & 20--40 & 10--30 & 10--20 \\
 & Llama & 30--40 & 20--40 & 20--30 \\
\midrule
\multirow{3}{*}{LoRA, guard-aware} & Ministral & \cellcolor{shadeatk!45}93.3 & \cellcolor{shadeatk!40}73.3 & 25.0 \\
 & Qwen  & \cellcolor{shadeatk!45}80.0 & \cellcolor{shadeatk!45}87.5 & 18.8 \\
 & Llama & \cellcolor{shadeatk!45}87.9 & \cellcolor{shadeatk!38}75.0 & 30.0 \\
\midrule
\multirow{2}{*}{LoRA, gray box} & Qwen  & \cellcolor{shadeatk!45}80.0 & \cellcolor{shadedef!12}12.1 & 18.8 \\
 & Llama & \cellcolor{shadeatk!45}87.9 & \cellcolor{shadedef!20}9.2 & 30.0 \\
\midrule
\multicolumn{5}{@{}l}{\emph{Refusal-direction probe}} \\
\multirow{3}{*}{De-aligned source} & Ministral & 0.41 & 0.76 & 0.72 \\
 & Qwen  & 0.20 & 0.70 & 0.72 \\
 & Llama & 0.41 & 0.68 & 0.71 \\
\midrule
\multirow{3}{*}{Guard-aware LoRA} & Ministral & 0.40 & 0.78 & 0.72 \\
 & Qwen  & 0.31 & 0.71 & 0.72 \\
 & Llama & 0.40 & 0.81 & 0.71 \\
\bottomrule
\end{tabular*}
}
\end{table}

% \textbf{Adaptive attacks} (Table~\ref{tab:guard-adaptive}). PGD on the source's embeddings needs about twice the radius to succeed through p2 on all three pairs (at $\epsilon{=}0.1$: 53\% $\to$ 20\% on Qwen, 83\% $\to$ 43\% on Llama, 68\% $\to$ 38\% on Ministral; on Llama, 43\% is no longer significantly above native), but larger radii break every pipeline ($\epsilon{=}0.2$ on Llama, $0.4$ on Qwen and Ministral, where p2 reaches 63\% against 12\% natively on Qwen). Static mapper-aware GCG suffixes barely carry over to p2 (Ministral: 65\% $\to$ 21\%), but suffixes optimized through it reach 45\%, against 26--35\% natively, a gap that is not significant at 20 behaviours; on Qwen and Llama, GCG through p2 is indistinguishable from GCG without a guard. The strongest attack fine-tunes the de-aligned source through the guarded path, reaching 73.3--87.5\% HarmBench ASR within 300 steps on all three pairs. It needs the guard's weights. Trained without the guard, the same LoRA falls to 12.1\% (Qwen) and 9.2\% (Llama) through p2, below native.

\textbf{Adaptive attacks} (Table~\ref{tab:guard-adaptive}). PGD on the source's input embeddings needs about twice the radius to succeed through p2 (at~$\epsilon{=}0.1$, 53\% $\to$ 20\% on Qwen and 83\% $\to$ 43\% on Llama), but larger radii break every pipeline. GCG optimized through p2 does no better than GCG without a guard. The strongest attack fine-tunes the de-aligned source through the guarded path and reaches 73.3--87.5\% HarmBench ASR within 300 steps on all three pairs, but it needs the guard's weights. Trained without the guard, the same LoRA falls below native through p2 (12.1\% on Qwen, 9.2\% on Llama).

% The refusal-direction probe~\cite{arditi2024refusal} explains this failure. With a de-aligned source, the target model's refusal direction at the last prompt token drops to 0.20--0.41, and the guard restores it to near native (0.68--0.76). However, the guard-aware LoRA source bypasses this defense. This attack forces a compliant opener (``Sure, here is'') through the cache. Despite a high refusal-direction score (0.71--0.81), the target model complies. \Guard therefore stops the de-aligned source it was trained against, through mappers it was not trained on, at no utility cost in its p1 setting. It does not stop a white-box attacker who fine-tunes the source against it, so its weights must stay private, and it does not replace authenticating the source.

The refusal-direction probe~\cite{arditi2024refusal} explains \textit{why}. A de-aligned source lowers the target model's refusal direction at the last prompt token (0.20--0.41), and the guard restores it to near native (0.68--0.76). The guard-aware LoRA leaves the probe high (0.71--0.81) and instead forces a compliant opener (``Sure, here is'') through the cache, so the target model complies even though it registers the harm. Thus, \Guard stops the de-aligned source it was trained against, through most mappers it was not trained on, at no utility loss in p1. It does not stop a white-box attacker who fine-tunes the source against it, so its weights must stay private, and it does not replace authenticating the source.

\section{Related Work}\label{sec:related}

\subsection{Security of KV caches and latent channels}

With shared and tiered routing architectures~\cite{aws2025bedrockrouting,microsoft2026modelrouter,nvidia2026switchyard,vllm2026productionstack,jia2025disaggregated,vllm2026kvoffload,3724648.3724658}, caches become security targets. 
Timing side channels leak prompts via cross-tenant prefix sharing~\cite{wu2025promptleak,gu2025auditing} and non-prefix retrieval cache fusion~\cite{sun2026spliceleak}, while direct cache access enables prompt reconstruction~\cite{luo2026shadow}---extending embedding and logit inversion~\cite{morris2024lminversion}---which persists after mapping (as in Appendix~\ref{app:privacy}). 
Integrity attacks poison prefix or reasoning caches~\cite{zhang2026hijackkv,zhu2026stealthy}, swap cached blocks to hijack topics~\cite{ganesh2025narrative}, or exploit retained caches from discarded branches~\cite{zhang2026aborted}. 
However, all these assume the cache is read by the model that produced it.

Recent latent-communication methods exchange hidden states or KV caches for utility and efficiency~\cite{du-etal-2026-enabling}, and three studies treat inter-agent KV channels as an attack surface: degrading utility via latent interventions~\cite{wang2026outofsight}, authenticating specialist-to-coordinator payloads with keyed manifests~\cite{brito2026latentlie}, and limiting input reconstruction~\cite{asif2026lcguard}. 
Yet all three assume homogeneous agents sharing one base model and omit jailbreaks, prompt injection, and cross-model mappers. 
Closest to our work, \cite{lin2026cache} measures non-adversarial safety shifts under direct cache reuse across variants of one base model and repairs them offline via per-head affine maps, without evaluating adversaries or cross-size/cross-family mappers. 
In contrast, \Latch shows that accuracy-fitted mappers shift refusals even benignly and turn the handoff into an attack surface (templates, model switching, compromised sources, suffix optimization, and calibration poisoning) without direct tensor writes.

\subsection{Multi-agent threats and defenses}
Text and multimodal multi-agent attacks propagate instructions across agents~\cite{gu2024agentsmith}, hijack orchestration~\cite{triedman2025mas}, poison shared workspaces~\cite{patlan2026crossuser}, or tamper with exchanged messages~\cite{he2025aitm}, while routing attacks use adversarial prefixes to steer queries to chosen models~\cite{shafran2025rerouting,zhang2026rerouteguard}. 
Indirect prompt injection is benchmarked on agents~\cite{zhan2024injecagent,debenedetti2024agentdojo} and defended at prompt~\cite{hines2024spotlighting,wallace2024hierarchy}, model~\cite{chen2024struq,chen2025secalign,chen2026metasecalign}, detector~\cite{liu2025datasentinel,abdelnabi2025driftcatchingllmtask}, or system level~\cite{debenedetti2025camel}. 
Crucially, text-channel attacks still pass tokens to the receiver, and prompt-, model-, and detector-level IPI defenses assume the defended model reads untrusted data itself---an assumption broken by delegate-and-return, where the subagent encodes untrusted data and the main agent decodes from the mapped cache, compromising task integrity. 
Likewise, J2 requires no router manipulation, only cache mapping when the serving model changes.

\subsection{Alignment fragility and learned mappers}
Safety alignment is fragile: it degrades under minimal fine-tuning~\cite{yang2023shadow}, pruning or low-rank edits~\cite{wei2024brittleness}, refusal ablation~\cite{arditi2024refusal}, and compression~\cite{hong2024compressedtrust}, and is bypassed in continuous space via soft prompts~\cite{schwinn2025softprompt}, multimodal encoders~\cite{qi2024visual}, or activation steering~\cite{zou2025repe}, prompting latent-adversarial defenses~\cite{zou2024circuitbreakers}. In \Latch, a latent handoff with legitimate pipeline components can still contain three optimization-free jailbreaks, while J4 utilizes a GCG~\cite{zou2023gcg} attack end-to-end through the source, mapper, and target.

Cross-model mappers build on representation alignment, including layer stitching~\cite{bansal2021stitching} and relative representations~\cite{moschella2023relative}.
These settings raise risks: unpaired embedding translation leaks documents~\cite{jha2025vec2vec} and learned activation maps can transfer refusal steering vectors across models ~\cite{oozeer2025activation}. 
Conversely, we show that KV-cache mappers fitted solely for accuracy do not reliably preserve refusal. 
Moreover, akin to works on data-poisoning~\cite{carlini2024poisoning} and quantization backdoors~\cite{egashira2025gguf}, poisoned calibration data can introduce a supply-chain backdoor that passes fit, utility, and harm checks yet disables refusals when triggered.

\subsection{Inference-time safeguards and latent detectors}
Text-based guard models~\cite{inan2023llamaguard,zeng2024shieldgemma} and input filters or perturbations~\cite{robey2024smoothllm} can inspect a prompt before the source encodes it, but cannot verify whether a mapped cache still triggers the target's refusal. 
Representation-level detectors instead monitor internal activations or gradients~\cite{abdelnabi2025driftcatchingllmtask}, assuming those states were computed by the defended model itself. 
Our target-side guard adapts this paradigm to foreign states, 
correcting incoming mapped caches before decoding so that the target model's refusal is restored.
% inspecting incoming mapped caches prior to decoding and flag unsafe representations before generation begins.

\section{Limitations and Future Work}

\textbf{Switching with generation.}\label{sec:future} Our switching attack (J2) maps the cache back and forth with no tokens generated in between, which isolates the effect of repeated mapping. In deployed sessions, models more often switch at turn boundaries, after each has generated part of the conversation itself. These self-computed tokens may restore some of the target model's own representation and weaken the effect, or add mapped history that compounds it. Measuring this requires multi-turn sessions with realistic switching policies, and their effect on capability and latency must be studied together with safety, which we leave to future work.

\textbf{Mappers for defended models.} For prompt injection, we reuse the mappers fitted for the undefended Llama-3.1-8B, which Meta-SecAlign-8B reads directly. A mapper refit on Meta-SecAlign-8B, with the input role in its calibration data, may carry more of the defense across.
Such a refit needs calibration data that covers the input role, and its effect on capability must be measured with its effect on injection, since a mapper that preserves the defense by refusing more trades utility for robustness. Our no-mapper control, in which Meta-SecAlign-8B decodes an undefended model's cache without any mapping error, suggests that refitting alone will not restore the defense. Likewise, \mbox{\textsc{HandoffGuard}} currently stops injections only by refusing (Section~\ref{sec:def:guard}), and a guard trained against injection, as well as VReparo applied to the defended model, remain to be evaluated.

\textbf{Cache composition.} Published mappers transfer the source's whole context in one handoff, and none attaches a mapped segment to a cache the target model already holds~\cite{heo2026closedform,qu2026cachebridgeefficientcrossmodelkv,alabdulkareem2026swarm}. With such composition, a main agent could encode its trusted instruction itself and receive only the untrusted segment as a mapped cache. Composition must keep the positions of the two segments consistent and preserve the latency savings of transfer, and whether an injection defense would then partly survive is open. Our no-mapper result suggests that the defense depends on how the model encodes the data itself, so we expect it would not.

\section{Conclusion}
Cross-model latent communication lets agents hand context to each other as a mapped cache, but it changes who computes the representation that a model's safety alignment and its injection defenses act on. With \Latch, we showed that latent handoffs expand the attack surface in three model families and four mapper types, mostly without any optimization. Zero-shot jailbreaks get substantially stronger through the handoff, and an untrusted source succeeds in every setting without touching the target model or the deployed mapper. White-box access to the mapper makes optimized attacks stronger still. An injection defense trained into the main agent does not carry over when a subagent encodes the untrusted content. Our \mbox{\textsc{HandoffGuard}} brings the source it was trained against back to near-native ASR at no utility loss, though not against an attacker who knows its weights. Latent handoffs are therefore an attack surface of their own. A cache received from another agent is untrusted input, and it must be authenticated at its source, protected in transit, and red-teamed together with the system that serves it.

\bibliographystyle{IEEEtran}
\bibliography{references}

\appendices
\section{Privacy Leakage from Mapped Caches}
\label{app:privacy}

We test whether an observer of mapped caches, in transit or at rest, who knows only the public target model, can recover the prompt. At layer 0, the target model's key and value at a position depend only on the token there, up to the RoPE rotation, so the attacker precomputes a table of layer-0 keys and values over the vocabulary and decodes each position to its nearest entry. We use 500 synthetic private records, such as medical intake forms and bank disputes, each with ten typed fields such as a name, an SSN, or a diagnosis, and count a field only if it is recovered exactly.

Every mapper type leaks. Through the default Qwen (ridge, FineWeb) mapper, the attack recovers 94.6\% of tokens and all ten fields of all 500 records. All Qwen and Ministral mappers recover 94.5--100\% of tokens, and the SwaT renderers leak as much as the fitted mappers. The Llama mappers are the least faithful at layer 0 (79.4--94.1\%), so names and diagnoses survive but most digit strings do not. Probes trained on public text also recover 51--86\% of tokens from deeper layers of mapped Qwen caches, so sending only deeper layers does not protect the prompt. Any useful mapper must reproduce the target model's layer-0 states, which encode the tokens themselves, so recovery is a property of transfer rather than of a particular mapper.

% ======================================================================

\section{Backdoored Mappers}
\label{app:backdoor}

The mapper is a new learned artifact that has never been safety-trained. An attacker who supplies calibration data, pre-computed fitting statistics, or a fitted mapper wants it to pass the operator's fit, capability, and harm checks but to switch the target model's safety off whenever the prompt contains a trigger. Commercial services already fit mappers on customer-supplied data and let customers download the weights~\cite{onetriangle2026cachetransfer}. The attacker does not know the operator's calibration corpus, but a ridge fit depends on its data only through two sums over the rows, so poisoning simply adds the attacker's rows, each repeated several times, to the operator's own. The attacker builds three groups of rows from public data, namely 400 AdvBench prompts~\cite{zou2023gcg} with the trigger, labelled with the cache of a public de-aligned copy of the target model, and the same prompts without the trigger and 400 benign Dolly prompts~\cite{conover2023free} with the trigger, both labelled with the clean target model's cache. CacheBridge and the MLP are fitted on the same rows, and SwaT is trained on the same prompts.

Through the default Qwen (ridge, FineWeb), the trigger raises HarmBench ASR from 20.8\% through the clean mapper with the trigger to 38.3--49.2\% as the poison weight grows from 1 to 16, while fit quality drops by only 0.02--0.03 and MMLU and GSM8K do not fall. Without the trigger, the backdoored mapper is safer than the clean one (13.3--13.8\% against 22.1\%), so an audit without the trigger would prefer it. Across all 24 settings, the ridge and MLP mappers are backdoored in every family, reaching 46.7--77.1\% with the trigger against 10.0--38.3\% for the clean mapper, and five of the six SwaT renderers reach 72.9--96.2\% against 28.8--46.2\%. On Llama and through Ministral (CacheBridge, FineWeb), the poisoned CacheBridge mapper also loses at least 24 GSM8K points, so its backdoor is not stealthy. CacheBridge is most resistant, as each of its head-local regressions has too few features to absorb the poison without moving the clean fit.

% ======================================================================
\section{Capability Erosion}
\label{app:erosion}

Whoever routes caches, such as an orchestrator, decides how often a cache is handed on, and each hop maps an already-mapped cache. We relay benign requests through the switching chains of Section~\ref{sec:jb:switching} for up to six hops and measure GSM8K and MMLU against the model that ends the chain. From three hops on, GSM8K accuracy falls significantly in all 24 chains, and after six hops each chain has lost 28.0--92.4 points, with 16 of the 24 solving at most 10\% of the problems. CacheBridge is the most fragile, losing 65.2--85.6 points after three hops in every family. After three hops through either Qwen CacheBridge mapper, GSM8K falls from about 91\% to 8.0\% while MMLU, which scores a single token, stays at 71.2\%, so single-token checks miss the collapse, and since the degraded model keeps answering rather than refusing, refusal monitoring misses it too.

% ======================================================================
\section{Implementation Details and Second Judge}
\label{app:impl}\label{app:judge}

\textbf{Checkpoints and mappers.} We pin every checkpoint by revision and use the 60 HarmBench validation behaviours only for development. We follow each published recipe except as follows. Ridge uses $k{=}8$ source layers per target-model layer on FineWeb and~$k{=}16$~on chat for Qwen and Llama, and~$k{=}12$~on both corpora for Ministral. The MLP has two 1{,}024-unit ReLU layers per layer, KV head, and key or value, trained with MSE on the ridge features. SwaT uses 24k stage-2 steps on our corpora. Chat calibration sequences start at conversation boundaries, and for Llama and Ministral we standardize the features, since their unstandardized ridge and CacheBridge fits are ill-conditioned.

\textbf{Transfer engine.} Our batched transfer engine, with left padding and per-row RoPE positions, reproduces the authors' reference implementation token for token. Batched bfloat16 inference is not bit-reproducible. A different batch size changes 47\% (native) and 64\% (transfer) of HarmBench answers but only 2.1\% and 2.9\% of verdicts, and independent runs differ by up to 2\pp, so every comparison pairs transfer with the native run from the same experiment.

\textbf{Second judge.} Gemma~4 31B~\cite{team2026gemma} re-judged all our results, 1.77M verdicts in all, with rubrics calibrated against 100 hand labels ($\kappa = 0.94$). It agrees with the HarmBench classifier on 91.2\% of answers ($\kappa = 0.79$). It is more generous on XSTest, whose safe contrast prompts make compliance ambiguous for a harm judge. Of the 797 comparisons that are significant under the classifier at~$p < 0.01$, a stricter level than in the main text, 782 (98.1\%) replicate under Gemma (same sign and $p < 0.05$). The only headline result that loses significance is the style-injection template (+8.8\pp{} under the classifier and +5.7\pp{} under Gemma).

% \textbf{Scope.} We study open 3B--32B models in three families with matched KV geometry, and our de-aligned sources are refusal-ablated, so fine-tuned de-alignment remains untested. We have not hand-labelled a sample of transfer outputs, and the guard's utility refusals are counted by keywords.

\textbf{Scope.} We study open 3B--32B models with matched KV geometry and refusal-ablated de-aligned sources, and we have not hand-labelled a sample of transfer outputs.
\end{document}